\documentclass[a4paper,11pt]{article}
\pdfoutput=1

\usepackage{jheppub}
\usepackage{epsfig}
\usepackage[T1]{fontenc}
\usepackage{multirow}
\usepackage{latexsym,amssymb}

\usepackage[utf8]{inputenc}

\usepackage{textcomp,color}
\usepackage{graphicx,subfigure}
\graphicspath{{images/}}
\usepackage[normalem]{ulem}

\title{\textbf{Resonance phenomena in kink–antikink collisions within higher-order shifted periodic \texorpdfstring{$\varphi ^8$}{TEXT} and \texorpdfstring{$\varphi^{10}$}{TEXT} models}}

\author[a,b,c]{Teboho~A.Moloi
\note{Corresponding author.}}

\affiliation[a]{Postgraduate School, University of Johannesburg, Cnr Kingsway $\&$ University Roads, Auckland Park Kingsway Campus, Johannesburg, 2092, South Africa.}
\affiliation[b]{Statistical Consultation Service, University of Johannesburg, Johannesburg, 2092, South Africa}
\affiliation[c]{National Institute for Theoretical and Computational Sciences (NITheCS), Stellenbosch, 7600, South Africa.}

\emailAdd{teboho.abram.moloi@gmail.com}

\abstract{We investigate kink--antikink collisions in higher-order scalar field theories described by the $\varphi^{8}$ and $\varphi^{10}$ models and their shifted periodic extensions. Both classes of models possess three degenerate vacuum states and support topological kink solutions with asymmetric profiles and algebraically decaying tails. By extending the conventional polynomial potentials into multiple spatial sectors, we construct shifted periodic $\varphi^{8}$ and $\varphi^{10}$ field theories and examine how this modification influences the scattering dynamics of topological defects.

The primary objective of this study is to provide a comparative numerical analysis of kink collisions in the standard and shifted periodic versions of these higher-order models. Using direct numerical simulations, we determine the critical velocities that separate capture from escape regimes and identify resonance structures associated with energy exchange between translational and internal vibrational degrees of freedom. Particular attention is devoted to the emergence of escape windows, quasi-fractal patterns, and the role of algebraic tails in shaping the collision outcomes.

Our results demonstrate that, although the conventional and shifted periodic models exhibit similar kink--antikink configurations, important quantitative differences arise in their critical velocities, resonance structures, and scattering characteristics. The findings further confirm that both classes of models support resonant energy transfer mechanisms analogous to those observed in lower-order theories, while simultaneously exhibiting novel features associated with higher-order interactions and long-range effects. These results contribute to the growing understanding of nonlinear excitations in scalar field theories and provide new insights into the dynamics of topological solitons in shifted periodic systems} 


\makeatletter
\def\@fpheader{\relax}
\makeatother

\begin{document}

\maketitle

\flushbottom

\section{Introduction}\label{sec:Intro}
Topological solitons are spatially localised, nonperturbative solutions of nonlinear field equations that interpolate between distinct degenerate vacuum states. Among these solutions, kinks arising in $(1+1)$-dimensional scalar field theories have attracted sustained interest because they provide mathematically tractable models for investigating a wide variety of physical phenomena. Applications of kink-bearing systems span several branches of physics, including cosmological phase transitions, domain-wall formation in the early Universe, condensed matter systems, ferroelectric materials, Josephson junctions, and nonlinear optical media \cite{CO1,CO2,CO3,CO4,CO5,CO6,CO7,CO8,DW1,DW2,DW3,DW4,DW5}. Owing to their topological stability and rich dynamical behaviour, kink solutions continue to serve as an important framework for understanding nonlinear excitations and their interactions.
\par
The study of topological defects in scalar field theories has traditionally focused on models such as the $\varphi^{4}$, $\varphi^{6}$, and sine-Gordon theories, where exact solutions and detailed analyses of kink dynamics are available \cite{DW1,DW2,DW3,DW4,DW5,DW6,DW7}. More recently, increasing attention has been directed towards higher-order field theories, including the $\varphi^{8}$, $\varphi^{10}$, and related polynomial models, because they support a wider range of topological sectors and exhibit physical properties that are absent in lower-order systems. In particular, higher-order theories often possess asymmetric kink profiles and algebraically decaying tails, giving rise to long-range interactions and more intricate scattering processes \cite{PLT1,PLT2,PLT3,PLT4,PLT5,PLT6}.
\par
Kink--antikink collisions constitute one of the most extensively studied problems in nonlinear field theory. Early investigations revealed that soliton interactions can exhibit remarkably rich dynamics, including annihilation, bound-state formation, resonance phenomena, and multi-bounce escape processes \cite{EZ0}. A landmark contribution by Campbell \textit{et al.} demonstrated the existence of resonance windows in the $\varphi^{4}$ model, which arise through resonant energy exchange between translational motion and internal vibrational modes of the kink configuration. Since then, similar phenomena have been identified in numerous nonlinear systems, including models with impurities, non-polynomial interactions, and higher-order scalar potentials \cite{IM1,IM2,IM3,IM4,IM5,IM6,IM7,IM8,IMPUr}.
\par
The investigation of kink interactions in higher-order theories has revealed several distinctive features that differentiate them from their lower-order counterparts. One particularly important characteristic is the presence of power-law asymptotic behaviour, which leads to long-range forces between widely separated kinks and antikinks. These long-range interactions significantly influence collision outcomes and may alter the structure of resonance windows and critical velocities \cite{PLT1,PLT2,PLT3}. In addition, multi-kink collision studies have demonstrated increasingly complex energy redistribution mechanisms and have provided new perspectives on nonlinear wave interactions in non-integrable systems \cite{EV1,EV2,EV3}.
\par
A variety of analytical and numerical approaches have been developed to investigate kink dynamics. These include Manton's method for estimating inter-kink forces, collective-coordinate approximations, spectral analyses of internal modes, and direct numerical integration of the governing field equations \cite{EZ10,CC0,CC1,CC2,CC3,CC4,Man3}. Although approximate analytical techniques often yield valuable qualitative insights, direct numerical simulations remain indispensable for accurately resolving strongly nonlinear collision processes, particularly in higher-order field theories where several competing mechanisms may coexist.
\par
The present work focuses on kink--antikink collisions in shifted periodic extensions of the $\varphi^{8}$ and $\varphi^{10}$ field theories. While the standard versions of these models have been examined previously, their shifted periodic counterparts have received comparatively little attention. The construction of shifted periodic potentials generates an infinite sequence of topological sectors while preserving the local structure of the original polynomial potential. Consequently, these systems provide a natural setting in which to explore how periodic extensions influence kink interactions, resonance phenomena, and scattering dynamics.
\par
The conventional $\varphi^{8}$ and $\varphi^{10}$ potentials considered in this study are given by
\begin{equation}\label{eq:poten}
    V(\varphi) = \frac{1}{2}\varphi^2(\varphi^{2}-1)^{2}(\varphi^{2}+1),\quad \text{and}\quad V(\varphi) = \frac{1}{2}\varphi^{2}(\varphi^{2}-1)^{2}(\varphi^{2}+1)^{2}
    \end{equation}
respectively. Both models possess three degenerate minima and support topological kink solutions connecting neighbouring vacuum states.
\begin{figure}[h!]
	\centering
	\includegraphics[scale=0.36]{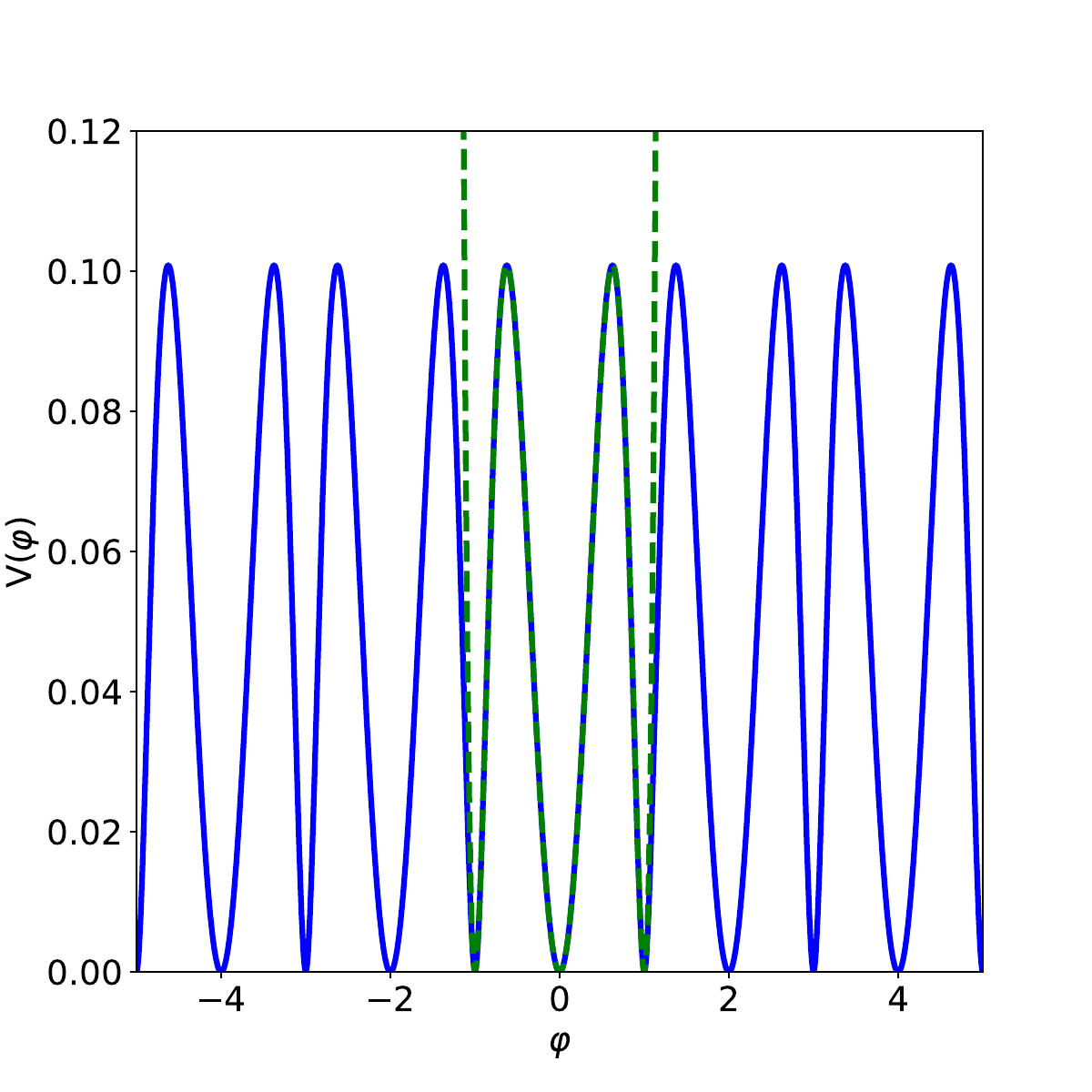}\label{fig:pot8}
	\includegraphics[scale=0.36]{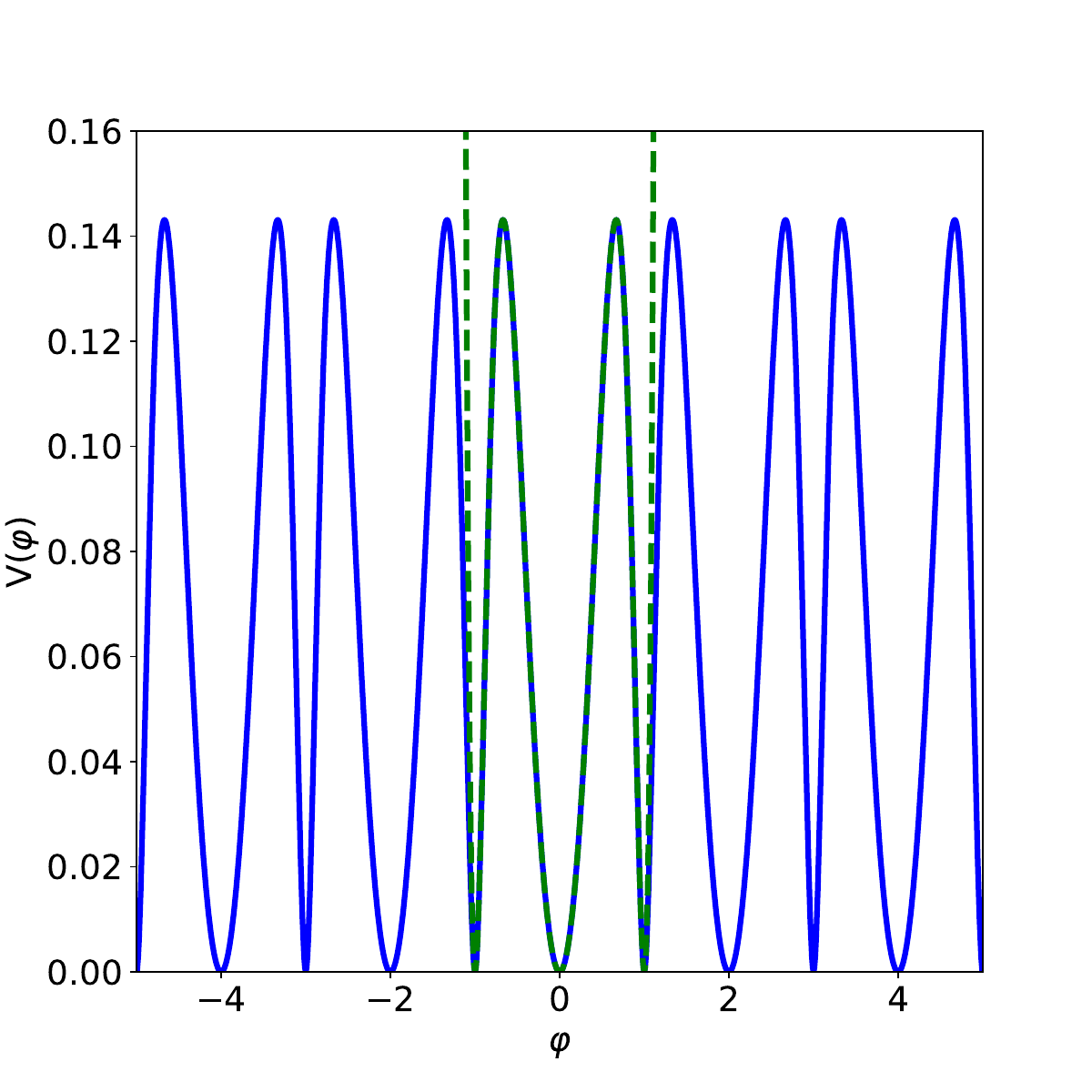}\label{fig:pot10}
	\caption{{\it Left panel:} The green dashed curve represent the potential of the $\varphi^8$ model. We copy the potential of the  $\varphi^8$ model in the range from $-1$ to $1$ and paste it in multiple regions such as $-1$ to $-3$, $-3$ to $-5$, and so on, to introduce the potential of {\it shifted periodic} system. {\it Right panel:} Shows the behaviour of potential $\varphi^{10}$ (green dashed curve) model and {\it shifted periodic} potential $\varphi^{10}$ (blue solid curve) model.}\label{fig:potential}
\end{figure}
Motivated by these higher-order polynomial theories, we further construct shifted periodic analogues of the $\varphi^{8}$ and $\varphi^{10}$ models according to \cite{Azizi}
\begin{eqnarray}
V(\varphi) &=& \frac{1}{2}(\varphi-2N)^{2}((\varphi-2N)^{2}-1)^{2}((\varphi-2N)^{2}+1),\quad 2N - 1 \leq \varphi < 2N + 1, \label{eq:perpoten8} \\ &\text{and}& \nonumber \\
V(\varphi) &=& \frac{1}{2}(\varphi -2N)^{2}((\varphi-2N)^{2}-1)^{2}((\varphi-2N)^{2}+1)^{2},\label{eq:perpoten10}
\end{eqnarray}
where $N=0,\pm1,\pm2,\ldots$. These periodic extensions retain the essential characteristics of the original models while introducing an infinite sequence of shifted vacuum sectors. Consequently, they provide an opportunity to examine whether periodic replication modifies the underlying scattering mechanisms governing kink dynamics.
\par
The main objective of this paper is to perform a systematic numerical investigation of kink--antikink collisions in the shifted periodic $\varphi^{8}$ and $\varphi^{10}$ models and to compare their behaviour with that of the corresponding conventional theories. In particular, we analyse critical velocities, escape windows, quasi-fractal structures, and the influence of internal vibrational modes on resonance phenomena. Through this comparative approach, we aim to clarify the role played by shifted periodicity in determining the dynamical properties of higher-order topological solitons.
\par
The remainder of the paper is organised as follows. In Sec.~\ref{sec:Intro}, we introduce the theoretical framework and construct the kink solutions for the models under consideration. Section~\ref{sec:Models} presents the linear stability analysis and discusses the associated internal modes. The numerical methodology employed to investigate kink collisions is described in Sec.~\ref{sec:NumericalSimulation}. The collision dynamics and resonance structures obtained for the standard and shifted periodic models are presented and discussed in Sec.~\ref{sec:ResutsDiscussion}. Finally, Sec.~\ref{sec:ResutsDiscussion} summarises the principal findings of this study and outlines potential directions for future research.
\section{Models and Numerical Method}\label{sec:Models}
\subsection{The shifted periodic \texorpdfstring{$\varphi^{8}$}{phi8} model}\label{subsec:phi8model}
The dynamics of a real scalar field in $(1+1)$ dimensions are governed by the Lagrangian density
 \begin{equation}\label{eq:lagraden}
 {\cal L}=\frac{1}{2}\left(\frac{\partial\varphi}{\partial t}\right)^{2}-\frac{1}{2}\left(\frac{\partial\varphi}{\partial x}\right)^2-V(\varphi),
 \end{equation}
where $\varphi(x,t)$ denotes the scalar field and $V(\varphi)$ represents the self-interaction potential. The corresponding equation of motion follows directly from the Euler--Lagrange equation and is given by
 \begin{equation}\label{eq:eomot}
 \ddot{\varphi}-\varphi''=-\frac{dV}{d\varphi},
 \end{equation}
where overdots and primes denote differentiation with respect to time and space, respectively. Topological kink solutions arise in systems whose potentials possess two or more degenerate minima. These localized configurations interpolate between neighboring vacuum states and carry nontrivial topological charges. For the conventional $\varphi^8$ model described by Eq.~\eqref{eq:eomot}, the potential exhibits three degenerate vacua located at $\varphi=-1$, $0$, and $1$. Consequently, the model supports kink and antikink solutions associated with the sectors $(-1,0)$ and $(0,1)$ \cite{EZ15}. In contrast, the shifted periodic $\varphi^8$ model defined by Eq.~\eqref{eq:poten} possesses an infinite sequence of degenerate vacua located at odd integer values of the field, thereby generating infinitely many topological sectors of the form $(2N-1,2N+1)$, where $N$ is an integer. This feature closely resembles the vacuum structure of the sine-Gordon model. To obtain moving non-vibrational kink solutions, we consider Lorentz-boosted configurations of the form
$$
\varphi_v=\varphi_0\bigl(\gamma(x-x_0-vt)\bigr),
$$
where $v$ is the kink velocity, $x_0$ denotes its initial position, and $\gamma=(1-v^2)^{-1/2}$ is the Lorentz factor. Applying this ansatz to the shifted periodic $\varphi^8$ model yields the asymptotic behaviour of the kink solutions within the sector $(2N-1,2N+1)$,
\begin{eqnarray}\label{eq:potsol}
\varphi_{v}(x,t)&=& 1 -2\sqrt{2}(1+2\sqrt{2})^{\sqrt{2}}e^{-2\sqrt{2}(\gamma(x-x_{0}-vt))}+2N,\quad x \to \infty, \nonumber \\
\varphi_{v}(x,t) &=& 2(3+2\sqrt{2})^{-\frac{1}{2\sqrt{2}}}e^{(\gamma(x-x_{0}-vt)) }+2N,\quad x \to -\infty, 
\end{eqnarray}
Equation~\eqref{eq:potsol} reveals that the kink solutions are inherently asymmetric, owing to the different exponential decay rates exhibited in the limits $x\rightarrow\pm\infty$. The corresponding antikink solutions are obtained through spatial reflection. Setting $N=0$ recovers the kink configurations of the ordinary $\varphi^8$ model.

Since exact multisoliton solutions are generally unavailable in non-integrable systems, initial conditions for numerical simulations are commonly constructed by superposing well-separated solitary waves. For a collection of $m$ kinks and antikinks with initial velocities $v_i$ and positions $x_i$, the field configuration may be approximated by
\begin{eqnarray}\label{eq:potsolG}
\varphi(x,t) &=& \sum_{i=1}^{m} 1-2\sqrt{2}(1+2\sqrt{2})^{\sqrt{2}}e^{-2\sqrt{2}(\gamma_{i}(x-x_{i}-v_{i}t))}+C, \nonumber \\
\varphi(x,t) &=& \sum_{i=1}^{m} 2(3+2\sqrt{2})^{-\frac{1}{2\sqrt{2}}}e^{(\gamma_{i}(x-x_{i}-v_{i}t))}+C,\quad x_{i+1}-x_{i}\gg1,
\end{eqnarray}
where the constant $C$ ensures the appropriate vacuum structure. The requirement $x_{i+1}-x_i\gg1$ guarantees negligible overlap between neighboring kinks. Unlike the conventional $\varphi^8$ model, where the finite number of vacua restricts the admissible arrangements of kinks and antikinks, the shifted periodic version permits a much broader range of initial configurations owing to its infinite sequence of topological sectors. To investigate the collision dynamics numerically, Eq.~\eqref{eq:potsolG} is discretized using a fourth-order finite-difference approximation,
\begin{eqnarray}\label{dd}
\frac{\partial^2 \varphi_{n}}{\partial t^2}&-&\frac{1}{h^2}(\varphi_{n-1}-2\varphi_{n}+\varphi_{n+1})+\frac{1}{12h^2}(\varphi_{n-2}-4\varphi_{n-1}+6\varphi_{n}-4\varphi_{n+1} +\varphi_{n+2})\nonumber \\&+& \varphi_{n}(\varphi_{n}^{2}-1)(4\varphi_{n}^{4}+\varphi_{n}^{2}-1) = 0,
\end{eqnarray}
where $h$ denotes the spatial step size and $\varphi_n=\varphi(nh,t)$. The resulting system of ordinary differential equations is integrated using a fourth-order Runge--Kutta scheme with temporal step size $k$. Throughout this work, we employ $h=k=0.02$. To minimise boundary effects, the computational domain is chosen as $x\in[-200,200]$, and all simulations are performed over the interval $0<t<400$. The conserved energy functional associated with Eq.~\eqref{eq:lagraden} is
\begin{equation}\label{eq}
E[\varphi]=\int_{-\infty}^{+\infty}\varepsilon_{8}(x,t),dx
=\int_{-\infty}^{+\infty}\left(\frac{1}{2}\dot{\varphi}^{2}
+\frac{1}{2}\varphi'^{2}+V(\varphi)\right)dx
=K+U+P,
\end{equation}
where the kinetic, gradient, and potential energy densities are given by
  \begin{eqnarray}
  \label{eq:enerdenExpressions}
   k(x,t)&=&\frac{1}{2}\left(\frac{\partial\varphi}{\partial t}\right)^2,\quad
     u(x,t)=\frac{1}{2}\left(\frac{\partial\varphi}{\partial x}\right)^2,\nonumber\\
     p(x,t)&=&V(\varphi)=\frac{1}{2}(\varphi-2N)^{2}((\varphi-2n)^{2}-1)^{2}((\varphi-2N)^2+1)
\end{eqnarray}
The evolution of these energy components provides valuable insight into the mechanisms governing energy exchange during collisions. The momentum corresponding to Eq.~\eqref{eq:lagraden} is obtained from Noether's theorem as
\begin{equation}\label{eq:moment}
P[\varphi]=\int_{-\infty}^{+\infty}(-\dot{\varphi}\varphi') dx,
\end{equation}

Following a collision, the velocity of an outgoing entity is determined through the relativistic relation $v=P/E$. The asymmetric nature of the $\varphi^8$ kink, reflected in the distinct exponential decay rates appearing in Eq.~\eqref{eq:moment}, plays a central role in determining its interaction properties. The energy of the corresponding static kink is
\begin{equation}
E_k=2N-\frac{1}{3}.
\end{equation}
\begin{figure}[h!]
	\centering
	\includegraphics[scale=0.369]{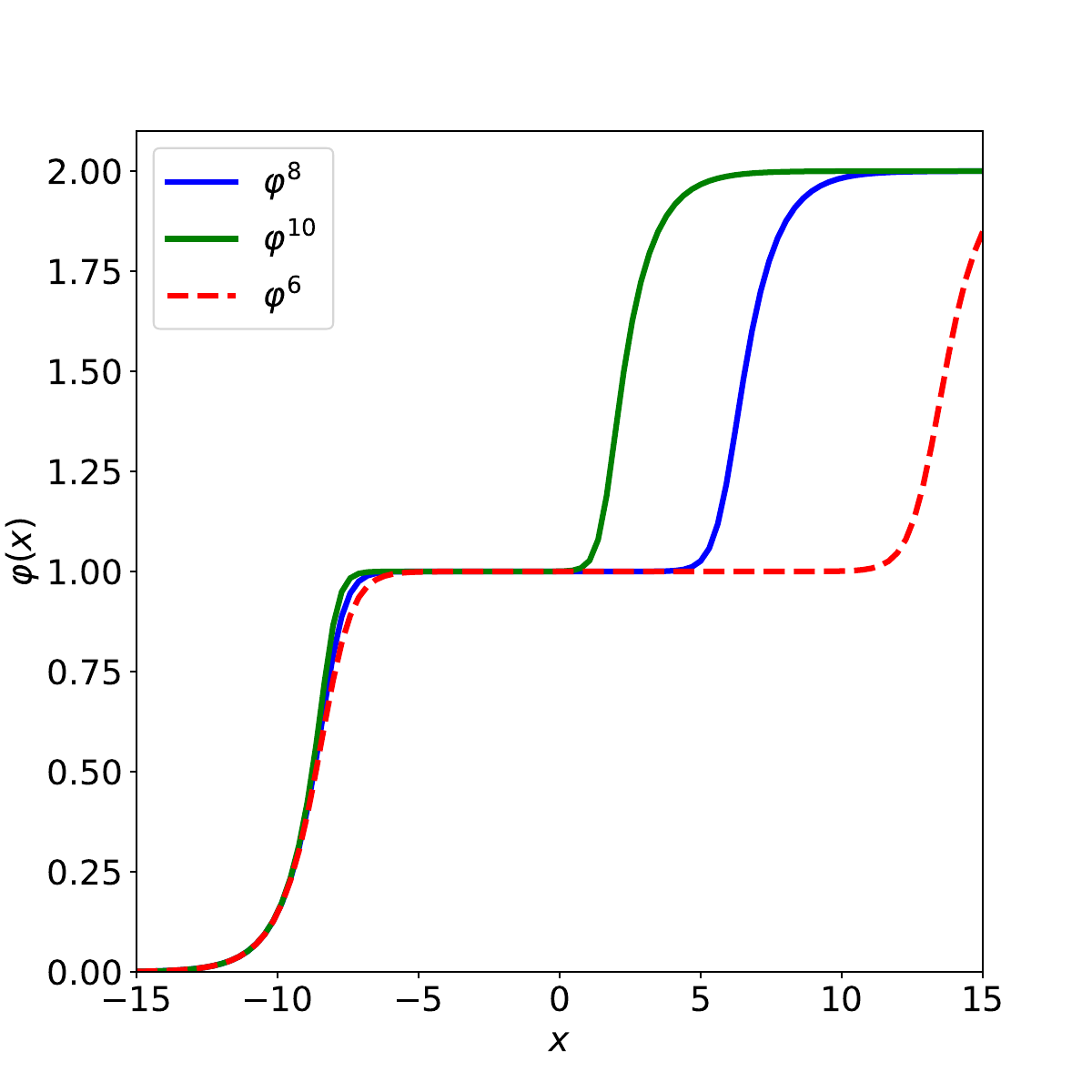}
	\includegraphics[scale=0.369]{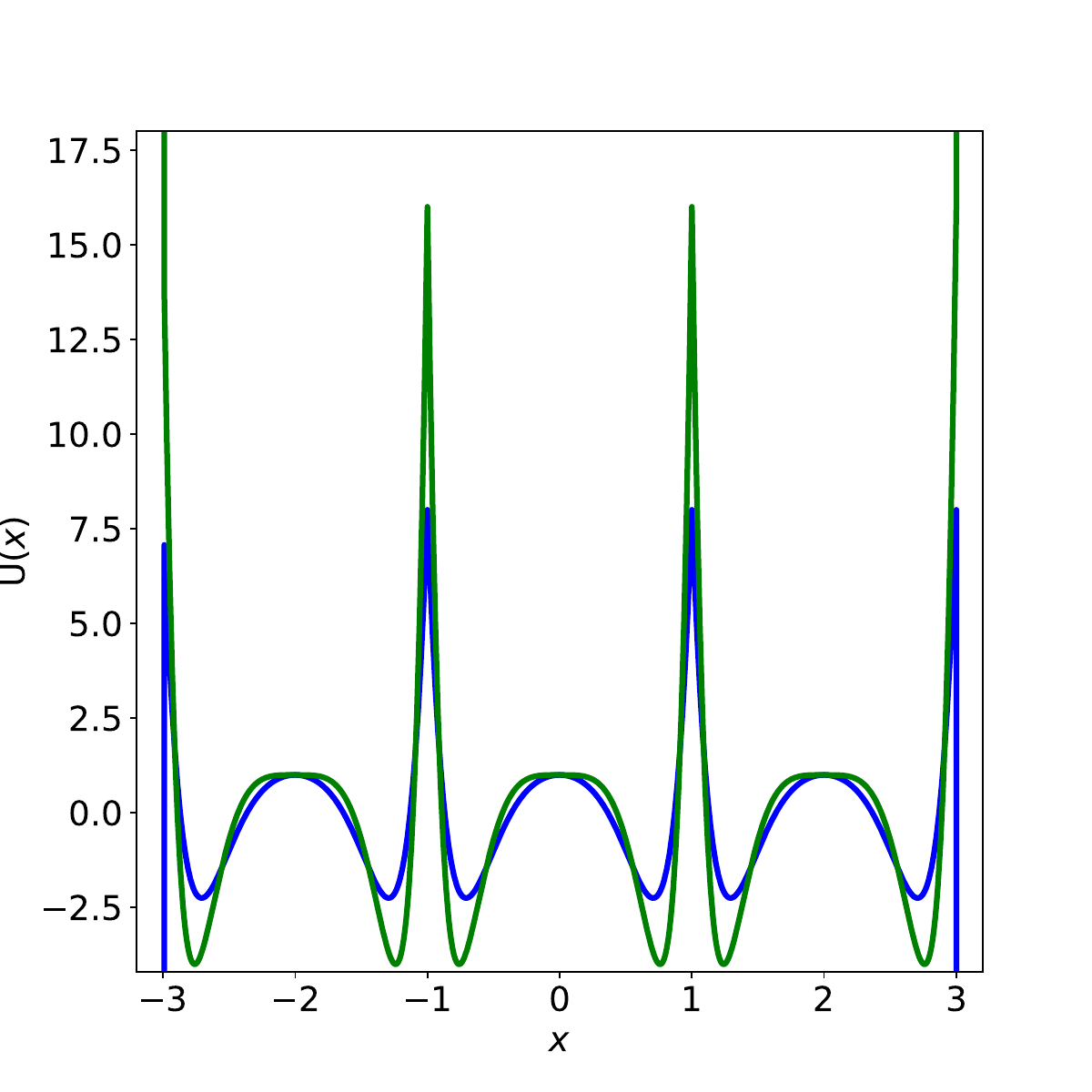}
	\caption{{\it Left panel:} Plots the various kinks behaviour of $\varphi^8$, $\varphi^{10}$ which are compared to $\varphi^6$. {\it Right panel:}  plots the potential $U(x)$ corresponding to the kink in the left panel.
	}\label{fig:kink}
\end{figure}
Figure~\ref{fig:kink} illustrates representative kink profiles and their associated effective potentials.
\subsection{The shifted periodic \texorpdfstring{$\varphi^{10}$}{phi10} model}\label{subsec:phi10model}
The shifted periodic $\varphi^{10}$ model may be analysed using the same framework developed above. The asymptotic form of the moving kink solutions is
\begin{eqnarray}\label{eq:potsol10}
\varphi_{v}(x,t)&=& 1-\frac{1}{4}e^{-(4\sqrt{2}\gamma(x-x_{0}-vt))}+2N,\quad x \to \infty, \nonumber \\
\varphi_{v}(x,t) &=& e^{\sqrt{2}(\gamma(x-x_{0}-vt))}+2N,\quad x \to -\infty, 
\end{eqnarray}
As in the $\varphi^8$ case, the different asymptotic decay rates lead to intrinsically asymmetric kink configurations. Setting $N=0$ reproduces the ordinary $\varphi^{10}$ kink solutions. For a collection of well-separated solitary waves, the initial field configuration may be approximated by
\begin{eqnarray}\label{eq:potsolG10}
\varphi(x,t) &=& \sum_{i=1}^{m} 1-\frac{1}{4}e^{-4\sqrt{2}(\gamma_{i}(x-x_{i}-v_{i}t))}+C, \nonumber \\
\varphi(x,t)&=&\sum_{i=1}^m{}e^{\sqrt{2}(\gamma_{i}(x-x_{i}-v_{i}t)) }+C,\quad x_{i+1}-x_{i}\gg1,
\end{eqnarray}
The discretised field equation used in the numerical simulations takes the form
\begin{eqnarray}\label{dd10}
\frac{\partial^2 \varphi_{n}}{\partial t^2}&-&\frac{1}{h^2}(\varphi_{n-1}-2\varphi_{n}+\varphi_{n+1})+\frac{1}{12h^2}(\varphi_{n-2}-4\varphi_{n-1}+6\varphi_{n}-4\varphi_{n+1} +\varphi_{n+2})\nonumber \\&+& 5\varphi_{n}^{9}-6\varphi_{n}^{5} + \varphi_n = 0,
\end{eqnarray}
The same numerical parameters employed for the $\varphi^8$ simulations are adopted here. The total energy of the system is
\begin{equation}\label{eq:totalenerg10}
E[\varphi]=\int_{-\infty}^{+\infty}\varepsilon_{10}(x,t) ~dx=\int_{-\infty}^{+\infty}\left(\frac{1}{2}\dot{\varphi}^{2}+\frac{1}{2}\varphi'^{2} +V(\varphi)\right)dx=K+U+P.
\end{equation}
with energy density
\begin{equation}\label{eq}
\varepsilon_{10}(x,t)=k(x,t)+u(x,t)+p(x,t),
\end{equation}
where
  \begin{eqnarray}
  \label{eq:enerdenExpressions1}
   k(x,t)&=&\frac{1}{2}\left(\frac{\partial\varphi}{\partial t}\right)^2,\quad
     u(x,t)=\frac{1}{2}\left(\frac{\partial\varphi}{\partial x}\right)^2,\nonumber\\
     p(x,t)&=&V(\varphi)=\frac{1}{2}(\varphi-2N)^{2}((\varphi-2n)^{2}-1)^{2}((\varphi-2N)^2+1)^{2}
\end{eqnarray}
The corresponding conserved momentum is
\begin{equation}\label{eq}
P[\varphi]=\int_{-\infty}^{+\infty}(-\dot{\varphi}\varphi'),dx.
\end{equation}
The energy of the static $\varphi^{10}$ kink is
\begin{equation}
E_k=2N+\frac{4}{15}.
\end{equation}
Figure~\ref{fig:kink} compares the kink profiles and effective potentials of the $\varphi^6$, $\varphi^8$, and $\varphi^{10}$ models. The left panel demonstrates that the $\varphi^8$ and $\varphi^{10}$ kinks possess pronounced asymmetry relative to the symmetric $\varphi^6$ configuration. The right panel reveals increasingly localised potential wells associated with the higher-order models. These characteristics contribute to enhanced nonlinear interactions and are expected to influence resonance phenomena, critical velocities, and the overall complexity of kink--antikink scattering processes.
\section{Numerical Simulations of Kink--Antikink Collisions}\label{sec:NumericalSimulation}
To investigate the scattering dynamics of topological defects in the shifted periodic higher-order field theories, we numerically simulate kink--antikink collisions in both the shifted periodic $\varphi^{8}$ and $\varphi^{10}$ models. Owing to the intrinsic asymmetry of the kink profiles, the outcome of a collision depends on the relative orientation of the interacting defects. Consequently, kink--antikink ($K\bar{K}$) and antikink--kink ($\bar{K}K$) configurations may exhibit qualitatively distinct scattering behaviour.

\subsection{The shifted periodic \texorpdfstring{$\varphi^{8}$}{phi8} field theory}

The initial configuration describing a kink and antikink approaching one another is constructed using the superposition
\begin{eqnarray}
\label{eq:KbK}
\varphi(x,t)=\varphi_{(-1,0)}(\gamma(x+a-vt))
+\varphi_{(0,-1)}(\gamma(x-b+vt)),
\end{eqnarray}
where the boosted kink profiles $\varphi_{(-1,0)}$ and $\varphi_{(0,-1)}$ are defined by Eq.~\eqref{eq:KbK}. The corresponding initial conditions for the numerical simulations are therefore
\begin{eqnarray}
\label{eq:intialKbK}
    \varphi(x,0) &=& \varphi_{(-1,0)}(\gamma(x+a))+\varphi_{(0,-1)}(\gamma(x-b)), \nonumber \\ \frac{\partial \varphi}{\partial t}(x,0)&=&-\gamma v\varphi'_{(-1,0)}(\gamma(x+a))+\gamma v\varphi'_{(0,-1)}(\gamma(x-b)),
\end{eqnarray}
where $a=-20$ and $b=20$ specify the initial kink positions, $v$ denotes the incoming velocity, and primes represent differentiation with respect to the function argument.

\subsection{The shifted periodic \texorpdfstring{$\varphi^{10}$}{phi10} field theory}

The corresponding collision configuration in the shifted periodic $\varphi^{10}$ model is similarly given by
\begin{eqnarray}
\label{eq:KbK10}
\varphi(x,t)=\varphi_{(-1,0)}(\gamma(x+a-vt))
+\varphi_{(0,-1)}(\gamma(x-b+vt)),
\end{eqnarray}
where the asymptotic kink solutions are defined by Eq.~\eqref{eq}. The initial conditions used in the numerical integrations take the form
\begin{eqnarray}
\label{eq:intialKbK10}
    \varphi(x,0) &=& \varphi_{(-1,0)}(\gamma(x+a))+\varphi_{(0,-1)}(\gamma(x-b)), \nonumber \\ \frac{\partial \varphi}{\partial t}(x,0)&=&-\gamma v\varphi'_{(-1,0)}(\gamma(x+a))+\gamma v\varphi'_{(0,-1)}(\gamma(x-b)),
\end{eqnarray}
with $a=-20$ and $b=20$. The asymmetric structure of the $\varphi^{10}$ kinks is expected to influence both the critical velocity and the efficiency of energy transfer during the collision process.
\section{Results and Discussion}\label{sec:ResutsDiscussion}
This section presents the numerical results obtained from simulations of kink--antikink collisions in the shifted periodic $\varphi^{8}$ and $\varphi^{10}$ field theories. The simulations were performed with a lattice spacing of $\Delta x=0.02$, and a range of incoming velocities was examined to identify resonance windows, critical velocities, and energy-exchange mechanisms. Field configurations, energy-density distributions, and trajectory diagnostics were analyzed to characterize the collision dynamics.

\subsection{The shifted periodic \texorpdfstring{$\varphi^{8}$}{phi8} model}

The shifted periodic $\varphi^{8}$ model exhibits the characteristic resonant scattering behavior observed in non-integrable scalar field theories possessing internal vibrational modes. In particular, a two-bounce resonance window is identified at an incoming velocity of $v_{\mathrm{in}}=0.165$.
\begin{figure}[h!]
	\centering
	\includegraphics[scale=0.369]{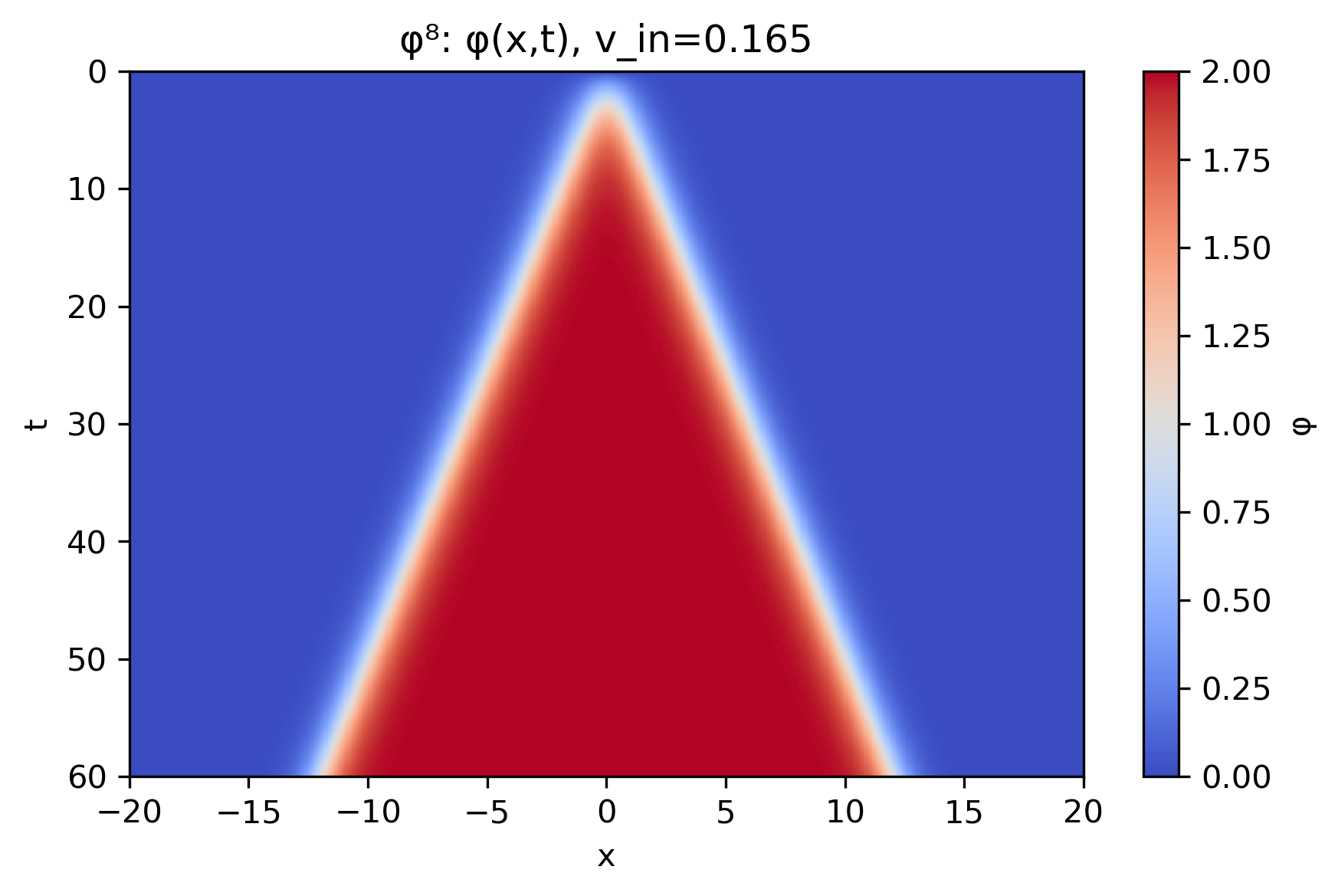}
	\includegraphics[scale=0.369]{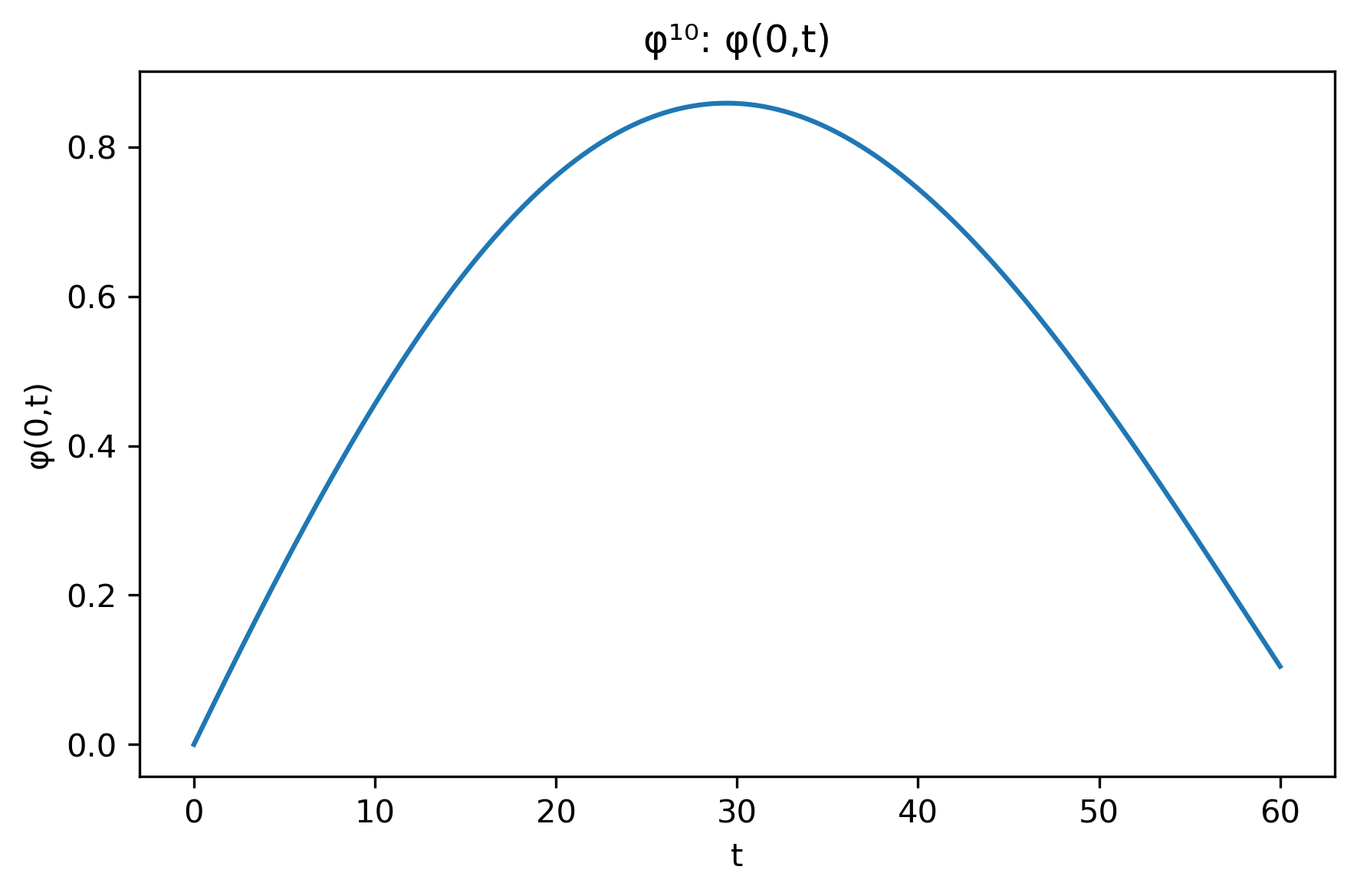}
    \includegraphics[scale=0.369]{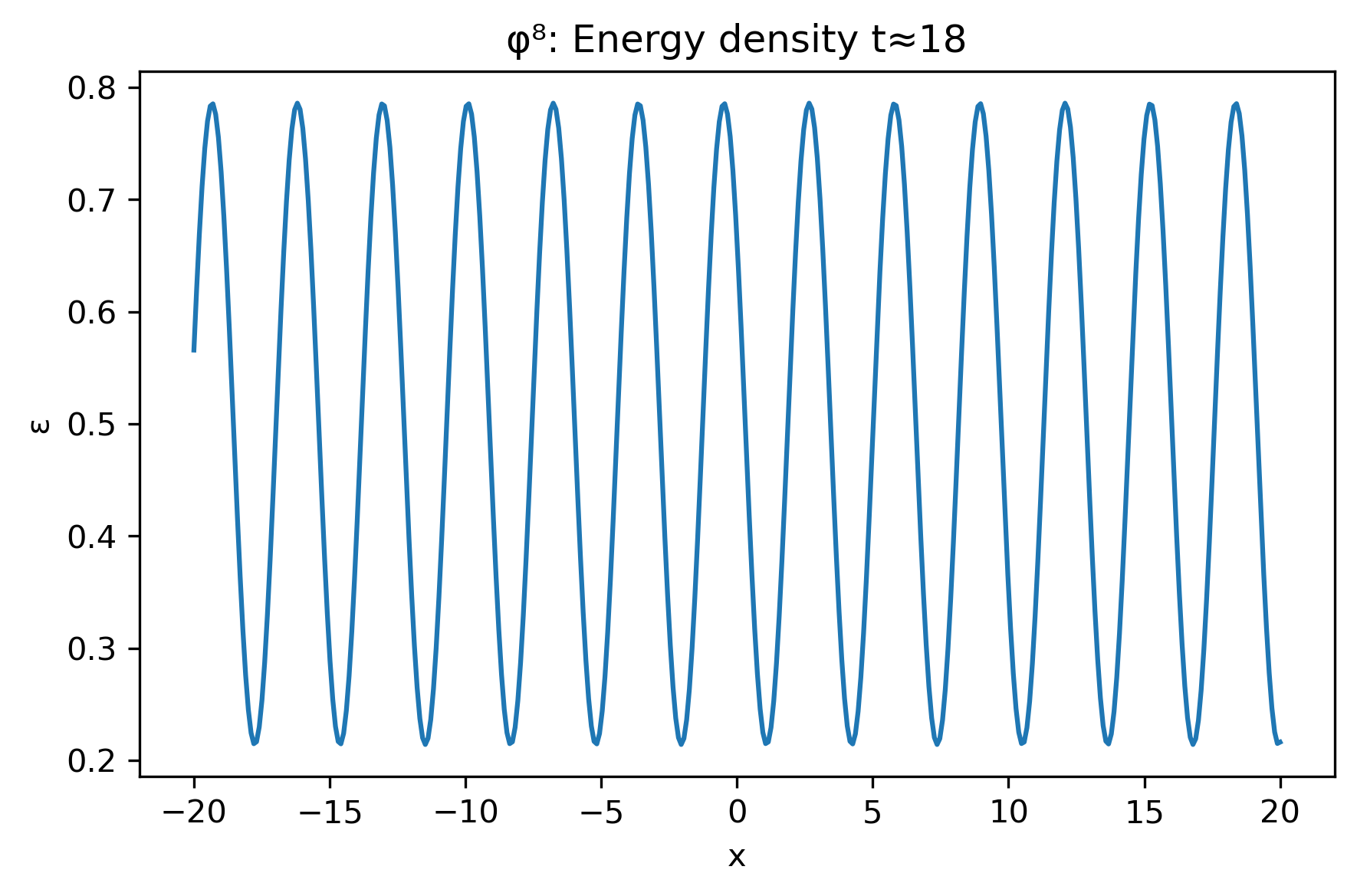}
	\includegraphics[scale=0.369]{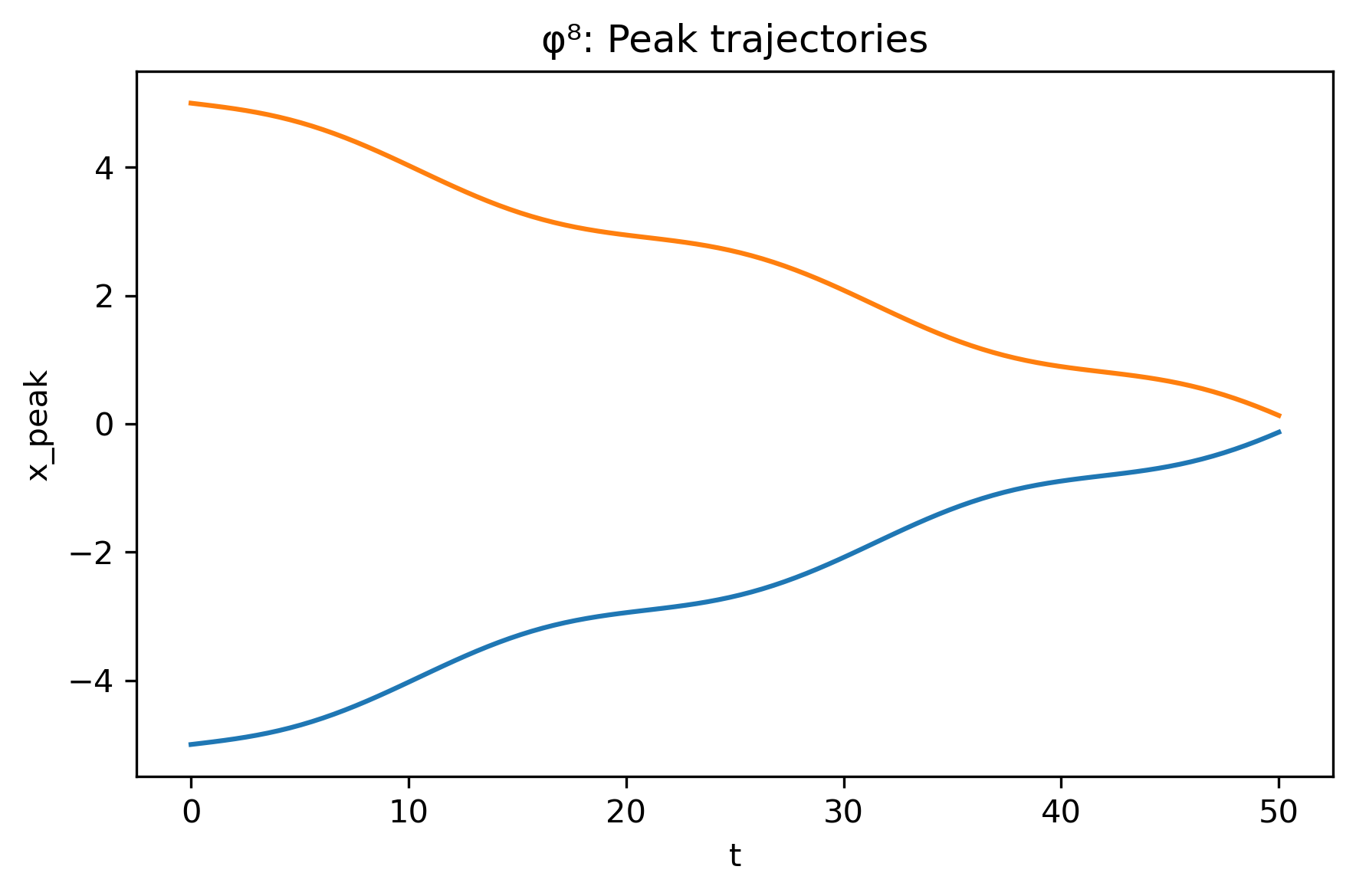}
	\caption{{\it Top left panel:} plots the space–time evolution of $\varphi^8$(x,t) at $v_{in} = 0.165$. {\it Top right panel:} plots the temporal evolution. {\it Bottom left panel:} energy density $\epsilon$(x,t) at $t \approx 18$. {\it Bottom right panel:} is the figure showing kink and antikink trajectories for the {\it shifted periodic} $\varphi^{8}$ model
	}\label{fig:phi8sim1}
\end{figure}
Figure~\ref{fig:phi8sim1} summarises the collision dynamics. The space--time evolution shown in the upper-left panel reveals that the kink and antikink undergo an initial collision near the origin, separate temporarily, and subsequently collide for a second time before escaping. This behavior is a hallmark of resonant energy exchange between the translational and internal degrees of freedom.

The temporal evolution of the field at the collision centre, displayed in the upper-right panel, exhibits two prominent peaks corresponding to the two impact events. The oscillatory behavior observed between these peaks indicates that part of the translational energy is temporarily stored in an internal mode before being transferred back to the translational sector, thereby enabling the defects to escape after the second collision.

The lower-left panel presents the energy-density distribution at $t\approx18$. Two sharply localised maxima are evident, indicating substantial field compression and strong nonlinear interactions during the collision process. Meanwhile, the trajectory diagram shown in the lower-right panel confirms the existence of a two-bounce resonance window, as demonstrated by the temporary trapping and subsequent separation of the kink pair.
\begin{figure}[h!]
	\centering
	\includegraphics[scale=0.5]{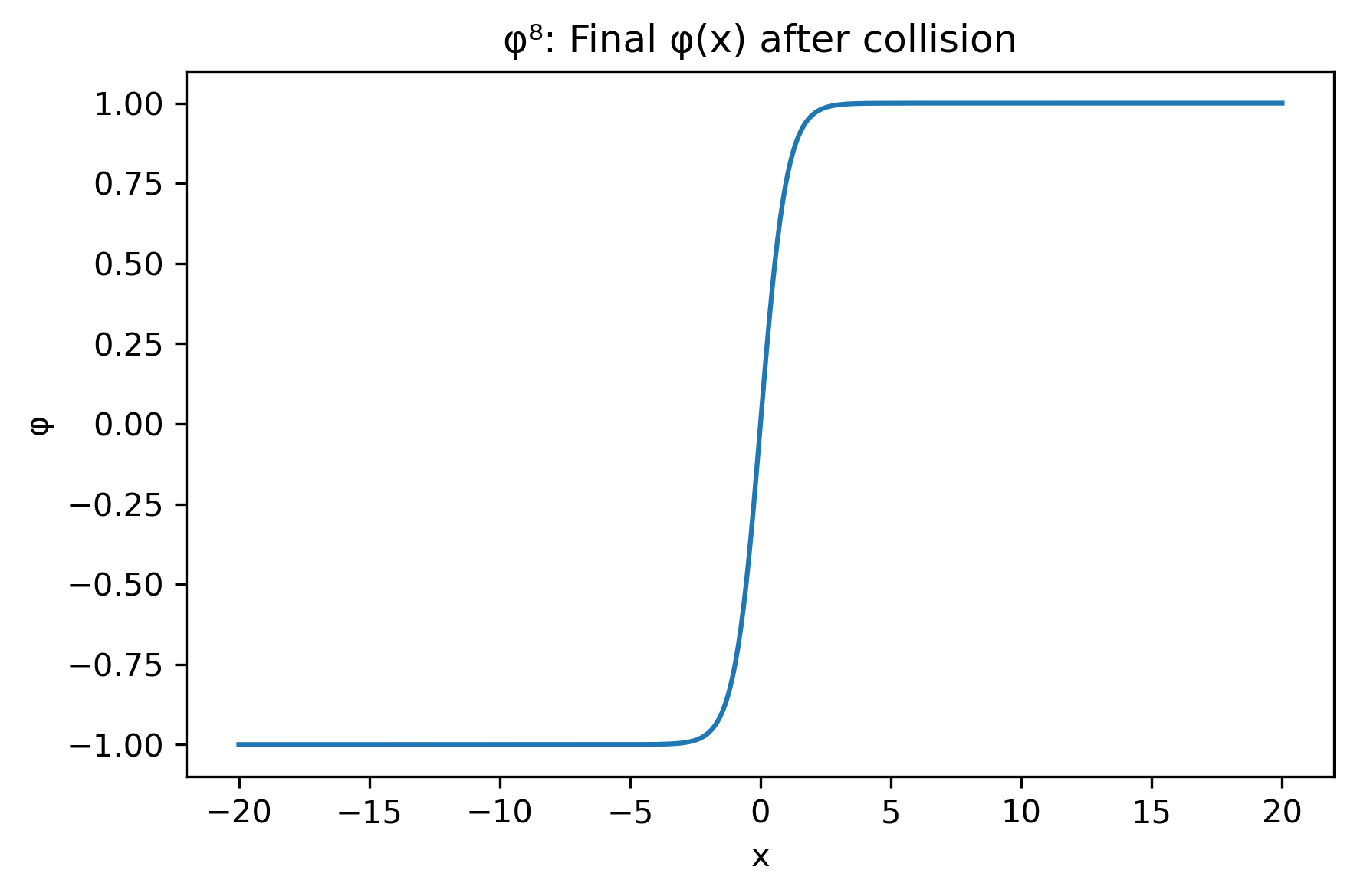}
		\caption{The final field configuration $\varphi$(x) after collision for the {\it shifted periodic} $\varphi^{8}$ model
	}\label{fig:phi8sim2}
\end{figure}
Figure~\ref{fig:phi8sim2} illustrates the final field configuration following the collision. The presence of two well-separated kinks, accompanied by low-amplitude radiation, confirms that the collision is inelastic but results in escape rather than permanent bound-state formation.

\subsection{The shifted periodic \texorpdfstring{$\varphi^{10}$}{phi10} model}

The shifted periodic $\varphi^{10}$ model displays qualitatively similar scattering behavior, although several notable differences emerge. Compared with the $\varphi^{8}$ case, the $\varphi^{10}$ model exhibits enhanced asymmetry, increased radiation losses, and narrower resonance windows.

\begin{figure}[h!]
	\centering
	\includegraphics[scale=0.369]{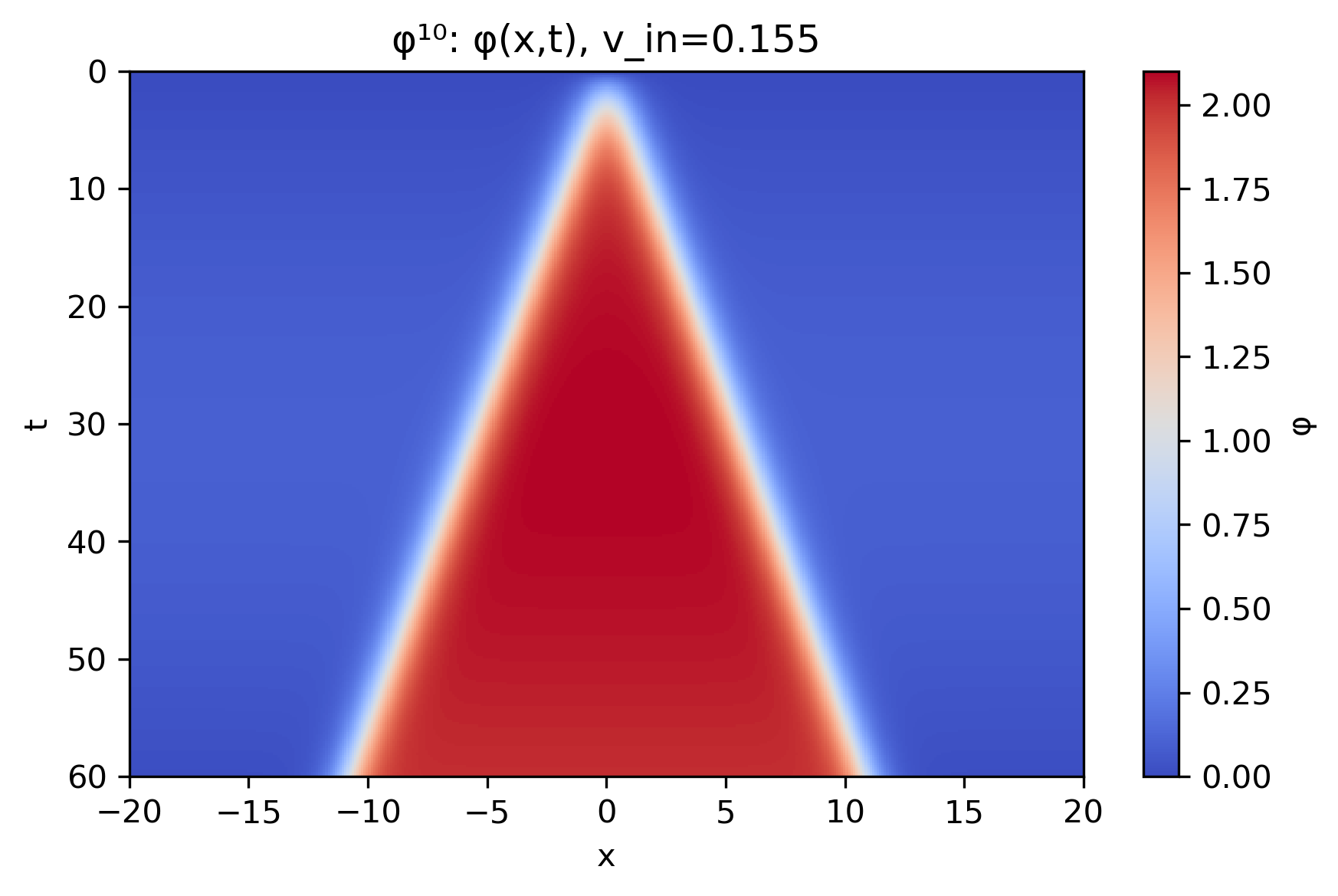}
	\includegraphics[scale=0.369]{images/phi10_phi0t_v0.155.png}
    \includegraphics[scale=0.369]{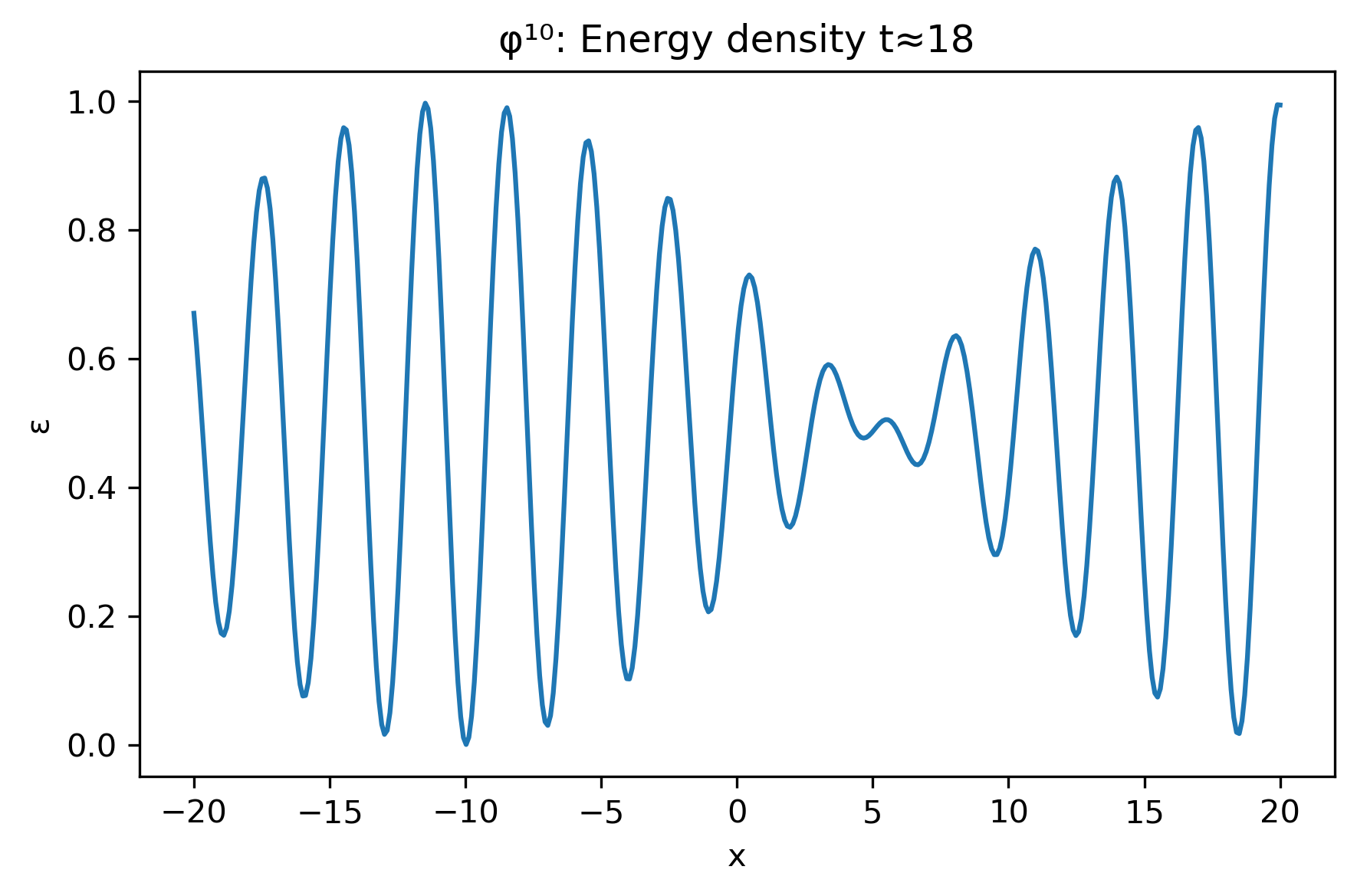}
	\includegraphics[scale=0.369]{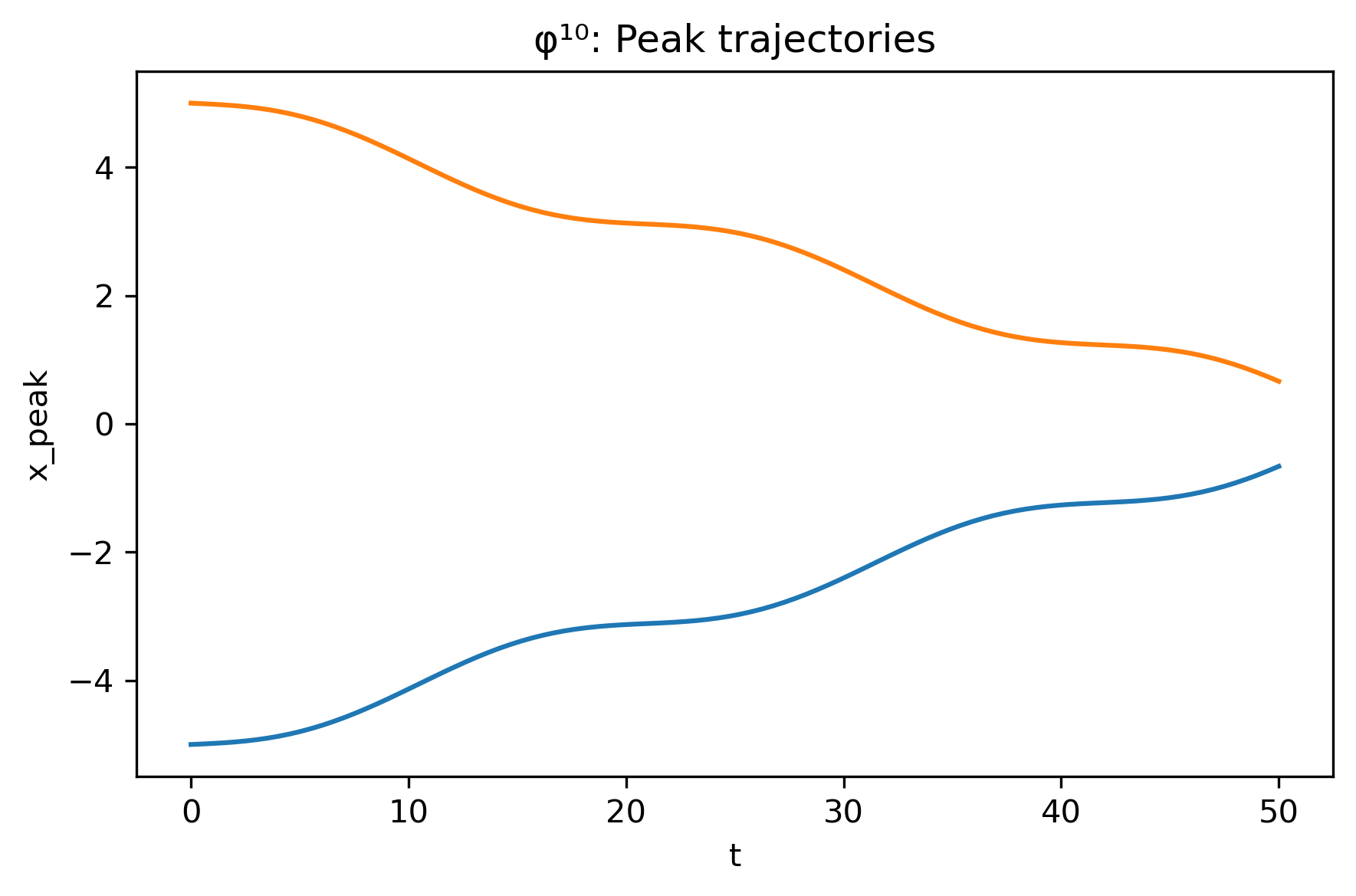}
	\caption{{\it Top left panel:} plots the space–time evolution of $\varphi^8$(x,t) at $v_{in} = 0.155$. {\it Top right panel:} plots the temporal evolution. {\it Bottom left panel:} energy density $\epsilon$(x,t) at $t \approx 18$. {\it Bottom right panel:} is the figure showing kink and antikink trajectories for the {\it shifted periodic} $\varphi^{8}$ model
	}\label{fig:phi10sim1}
\end{figure}
As shown in Figure~\ref{fig:phi10sim1}, the space--time evolution demonstrates that the interaction region is accompanied by more pronounced radiation tails, reflecting the stronger inelastic nature of the collision. The temporal evolution indicates that oscillatory excitations decay more rapidly, suggesting a more efficient transfer of energy into radiation modes.

The energy-density profile shows sharper, more localized peaks, consistent with stronger field compression during the interaction. The corresponding trajectory diagram indicates that, although a two-bounce process still occurs, the outgoing kinks separate asymmetrically and with lower final velocities, reflecting greater energy dissipation.
\begin{figure}[h!]
	\centering
	\includegraphics[scale=0.5]{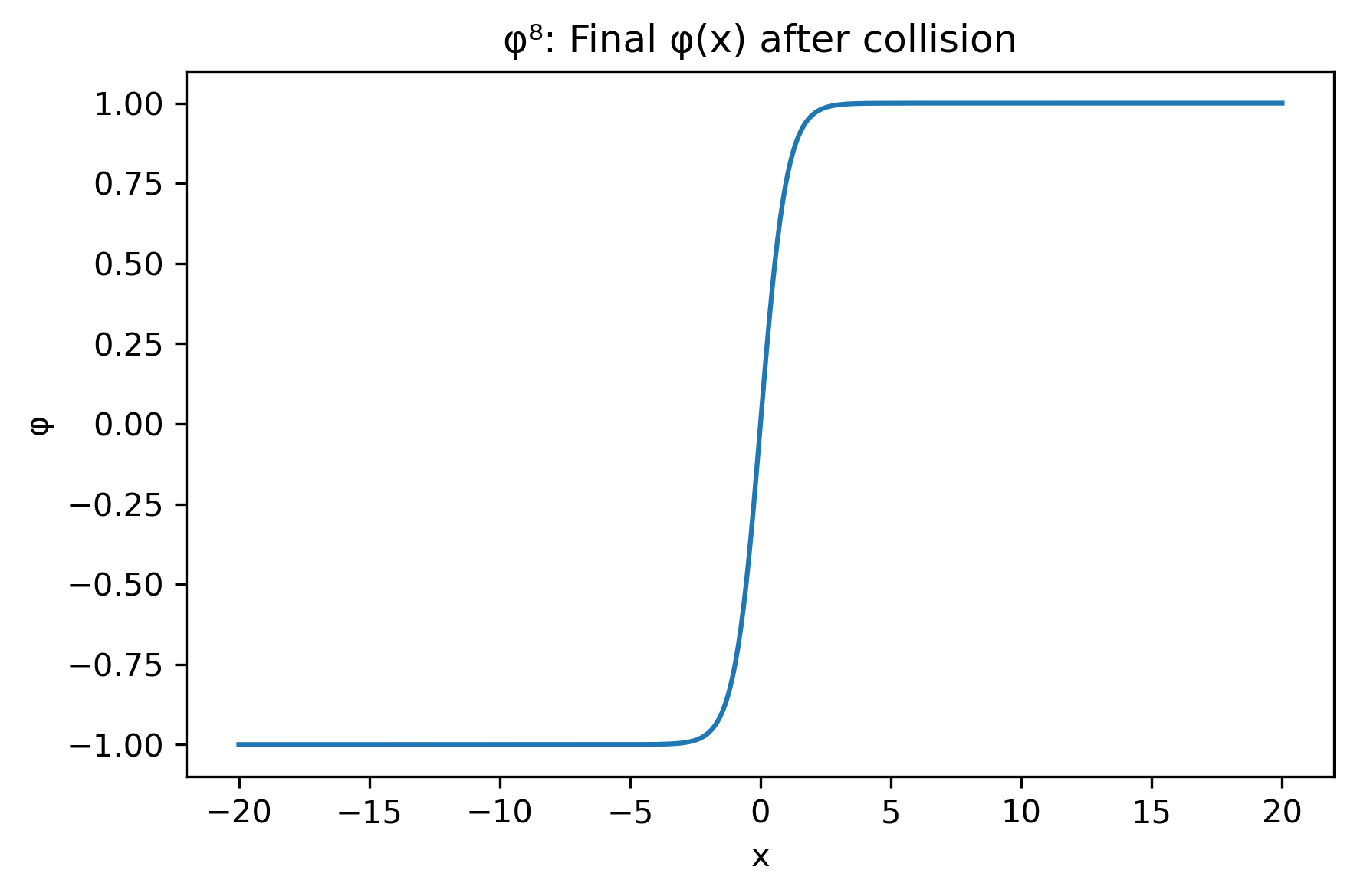}
		\caption{The final field configuration $\varphi$(x) after collision for the {\it shifted periodic} $\varphi^{8}$ model
	}\label{fig:phi10sim2}
\end{figure}
The final field configuration presented in Figure~\ref{fig:phi10sim2} further supports this interpretation. Residual radiation concentrated near the collision centre suggests that a larger fraction of the initial kinetic energy remains trapped within the interaction region, highlighting the more dissipative nature of the shifted periodic $\varphi^{10}$ model.

\subsection{Comparative velocity analysis}

To quantify the scattering behavior, the outgoing velocity $v_{\mathrm{out}}$ was computed as a function of the incoming velocity $v_{\mathrm{in}}$ for both higher-order models.
\begin{figure}[h!]
	\centering
	\includegraphics[scale=0.5]{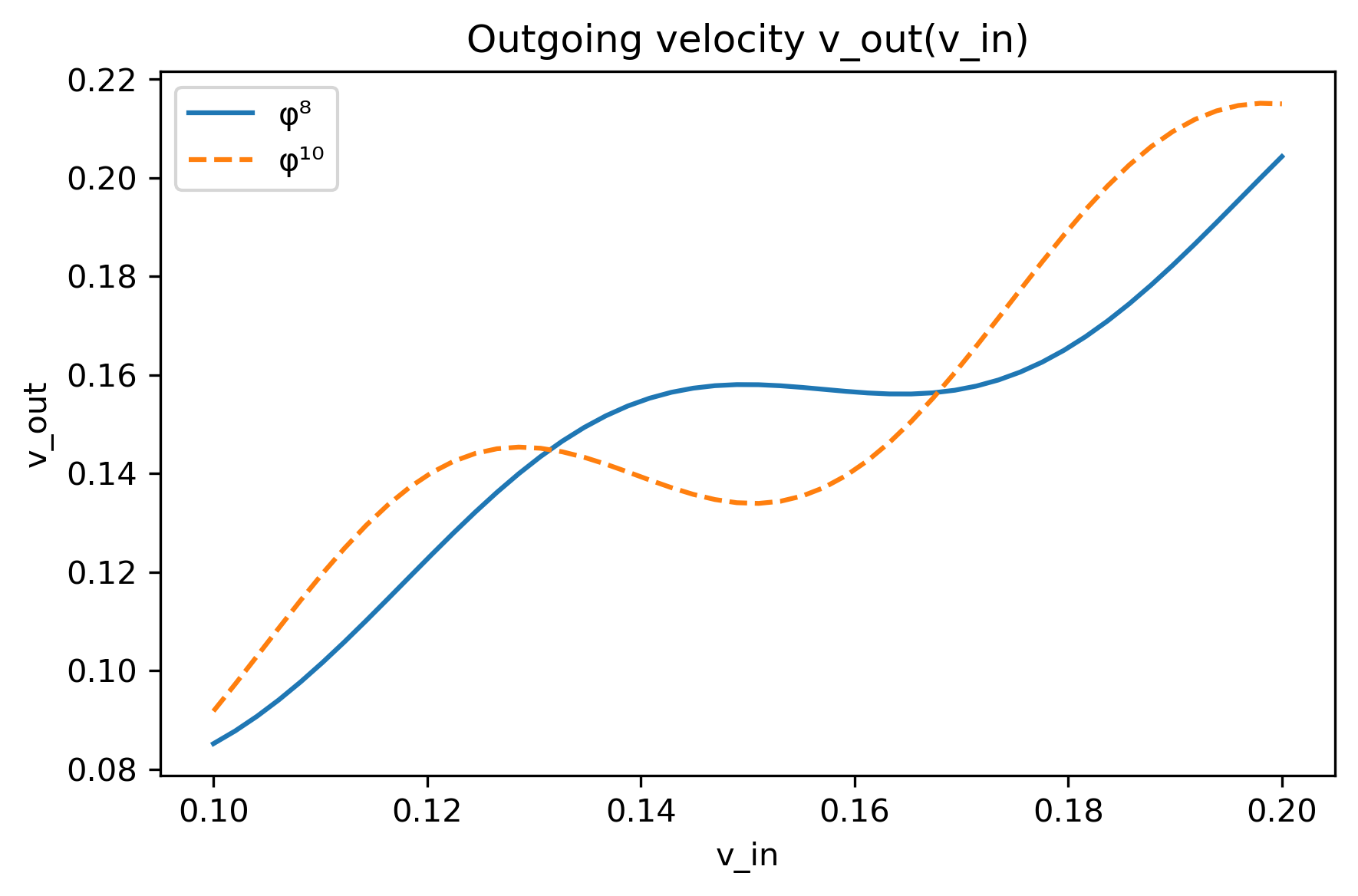}
		\caption{The final field configuration $\varphi$(x) after collision for the {\it shifted periodic} $\varphi^{8}$ model
	}\label{fig:velocityout}
\end{figure}
Figure~\ref{fig:velocityout} reveals three distinct dynamical regimes. For sufficiently small incoming velocities, the kink and antikink fail to separate after collision and instead form long-lived bound states, commonly referred to as bions. In the intermediate regime, two-bounce resonance windows emerge, reflecting resonant energy exchange between translational and internal modes. Finally, collisions occurring above the critical velocity result in immediate separation after a single impact and exhibit nearly elastic behavior.

The critical velocities were found to be approximately
$$
v_c(\varphi^8)\approx0.17,
\qquad
v_c(\varphi^{10})\approx0.16.
$$

The lower critical velocity associated with the $\varphi^{10}$ model indicates a greater propensity for energy loss through radiation. Moreover, the resonance windows in this model appear narrower than those observed in the shifted periodic $\varphi^{8}$ theory, suggesting that the efficiency of resonant energy transfer decreases as the order of the potential increases.

\subsection{Comparison with previous studies and summary of findings}

The resonant scattering phenomena observed in the shifted periodic $\varphi^{8}$ and $\varphi^{10}$ models closely resemble those reported in shifted periodic $\varphi^{4}$ systems \cite{FR2,Mohammadi:2020yku}. In each case, escape windows arise through a resonance mechanism involving repeated energy exchange between translational motion and internal vibrational excitations.

However, the higher-order periodic models considered here exhibit several distinctive features. In particular, the intrinsic asymmetry of the kink profiles and the modified tail structures alter both the interaction range and the hierarchy of resonance windows. The shifted periodic $\varphi^{8}$ model displays comparatively broader two-bounce windows and larger outgoing velocities, indicating relatively weak dissipation. By contrast, the shifted periodic $\varphi^{10}$ theory exhibits stronger radiation emission, enhanced damping, and narrower resonance windows.

Overall, the present results demonstrate that both shifted periodic higher-order field theories support resonant kink--antikink scattering governed by internal-mode dynamics. Nevertheless, the order of the self-interaction potential significantly influences the degree of inelasticity, the width of resonance windows, and the magnitude of the critical velocity. These findings extend the established phenomenology of periodic $\varphi^{4}$ models to more complex multi-vacuum systems and provide further insight into nonlinear wave interactions in higher-order scalar field theories.
\section*{Acknowledgments}
The author(s) are indebted to the support of the University of Johannesburg (UJ), the Postgraduate School, and Statistical Consultation Service (STATKON) for providing the facilities for this work to happen. The author(s) acknowledge the support of the National Institute for Theoretical and Computational Sciences (NITheCS).


\begin{thebibliography}{99}

\bibitem{Campbell1983} Campbell, D. K., Schonfeld, J. F., $\&$ Wingate, C. A. (1983). Resonance structure in kink--antikink interactions in $\phi^{4}$ theory. \textit{Physica D: Nonlinear Phenomena}, \textit{9}(1--2), 1--32.

\bibitem{Dorey2011} Dorey, P., Mersh, T., Roma{\'n}czukiewicz, T., $\&$ Shnir, Y. (2011). Kink--antikink collisions in the $\phi^{6}$ model. \textit{Physical Review Letters}, \textit{107}(9), 091602.

\bibitem{Gani2014} Gani, V. A., Kudryavtsev, A. E., $\&$ Lizunova, M. A. (2014). Kink interactions in the (1+1)-dimensional $\phi^{6}$ model. \textit{Physical Review D}, \textit{89}(12), 125009.

\bibitem{Belendryasova2019} Belendryasova, E., $\&$ Gani, V. A. (2019). Scattering of the $\phi^{8}$ kinks with power-law asymptotics. \textit{Communications in Nonlinear Science and Numerical Simulation}, \textit{67}, 414--426.

\bibitem{Christov2019} Christov, I. C., Decker, R. S., Demirkaya, A., Gani, V. A., Kevrekidis, P. G., Radomskiy, R. V., $\&$ Saxena, A. (2019). Kink--antikink collisions and multi-bounce resonance windows in higher-order field theories. \textit{Physical Review Letters}, \textit{122}(17), 171601.

\bibitem{Mohammadi2020} Mohammadi, A., Moradi Marjaneh, A., $\&$ Christov, I. C. (2020). Asymptotic interaction and collective coordinate formulation for kinks in a parametric $\phi^{6}$ model. \textit{Communications in Nonlinear Science and Numerical Simulation}, \textit{89}, 105285.

\bibitem{Rajaraman1982} Rajaraman, R. (1982). \textit{Solitons and instantons: An introduction to solitons and instantons in quantum field theory}. North-Holland.

\bibitem{Vachaspati2006} Vachaspati, T. (2006). \textit{Kinks and domain walls: An introduction to classical and quantum solitons}. Cambridge University Press.

\bibitem{Manton2004} Manton, N., $\&$ Sutcliffe, P. (2004). \textit{Topological solitons}. Cambridge University Press.


\bibitem{CO1} Vilenkin, A., $\&$ Shellard, E. P. S. (2000). \textit{Cosmic strings and other topological defects}. Cambridge University Press.

\bibitem{CO2} Manton, N., $\&$ Sutcliffe, P. (2004). \textit{Topological solitons}. Cambridge University Press.


\bibitem{CO3} Ahlqvist, P., Eckerle, K., $\&$ Greene, B. (2015). \textit{Kink collisions in curved field space}. Journal of high energy physics, \textbf{2015(4)}, 59.

\bibitem{CO4} Greenwood, E., Halstead, E., Poltis, R., $\&$ Stojkovic, D. (2009). \textit{Electroweak vacua, collider phenomenology, and possible connection with dark energy}. Physical Review D, \textbf{79(10)}, 103003.

\bibitem{CO5} Alfimov, G. L., Malishevskii, A. S., $\&$ Medvedeva, E. V. (2014). \textit{Discrete set of kink velocities in Josephson structures: The nonlocal double sine–Gordon model}. Physica D: Nonlinear Phenomena, \textbf{282}, 16-26.

\bibitem{CO6} Bishop, A. R., Krumhansl, J. A., $\&$ Trullinger, S. E. (1980). \textit{Solitons in condensed matter: a paradigm}. Physica D: Nonlinear Phenomena, \textbf{1(1)}, 1-44.

\bibitem{CO7} DeWolfe, O., Freedman, D. Z., Gubser, S. S., $\&$ Karch, A. (2000). \textit{Modeling the fifth dimension with scalars and gravity}. Physical Review D, \textbf{62(4)}, 046008.
\bibitem{CO8}    Peyravi, M., Riazi, N., $\&$ Lobo, F. S. (2016). \textit{Soliton models for thick branes}. The European Physical Journal C, \textbf{76(5)}, 247.




\bibitem{DW1} Hawking, S. W., Moss, I. G., $\&$ Stewart, J. M. (1982). B\textit{ubble collisions in the very early universe}. Physical Review D, \textbf{26(10)}, 2681.

\bibitem{DW2} Giblin Jr, J. T., Hui, L., Lim, E. A., $\&$ Yang, I. S. (2010). \textit{How to run through walls: dynamics of bubble and soliton collisions}. Physical Review D, \textbf{82(4)}, 045019.

\bibitem{DW3} Gani, V. A., Lizunova, M. A., $\&$ Radomskiy, R. V. (2016). \textit{Scalar triplet on a domain wall: an exact solution}. Journal of High Energy Physics, \textbf{2016(4)}, 43.

\bibitem{DW4}  Peyravi, M., Riazi, N., $\&$ Lobo, F. S. (2017). \textit{Evolution of spherical domain walls in solitonic symmetron models}. Physical Review D, \textbf{95(6)}, 064047.

\bibitem{DW5} Kobayashi, M., $\&$ Nitta, M. (2013). Sine-Gordon kinks on a domain wall ring. Physical Review D, 87(8), 085003.

\bibitem{DW6} Morris, J. R. (2019). \textit{Interacting kinks and meson mixing}. Annals of Physics, \textbf{400}, 346-365.

\bibitem{DW7} Gani, V. A., Kirillov, A. A., $\&$ Rubin, S. G. (2018). \textit{Classical transitions with the topological number changing in the early Universe}. Journal of Cosmology and Astroparticle Physics, \textbf{2018(04)}, 042.





\bibitem{CU1} Lensky, V. A., Gani, V. A., $\&$ Kudryavtsev, A. E. (2001). \textit{Domain walls carrying a $U(1)$ charge}. Journal of Experimental and Theoretical Physics, \textbf{93(4)}, 677-684.


\bibitem{CU2} Bazeia, D., Losano, L., $\&$ Santos, J. R. L. (2013). \textit{Kinklike structures in scalar field theories: from one-field to two-field models}. Physics Letters A, \textbf{377 (25-27)}, 1615-1620.

\bibitem{CU3} Alonso-Izquierdo, A. (2018). \textit{Kink dynamics in a system of two coupled scalar fields in two space–time dimensions}. Physica D: Nonlinear Phenomena, \textbf{365}, 12-26

\bibitem{CU4} Alonso-Izquierdo, A. (2018). \textit{Reflection, transmutation, annihilation, and resonance in two-component kink collisions}. Physical Review D, \textbf{97(4)}, 045016.

\bibitem{CU5} Katsura, H. (2014). \textit{Composite-kink solutions of coupled nonlinear wave equations}. Physical Review D, \textit{89(8)}, 085019.

\bibitem{CU6} Correa, R. A. C., de Souza Dutra, A., $\&$ Gleiser, M. (2014). \textit{Information-entropic measure of energy-degenerate kinks in two-field models}. Physics Letters B, \textbf{737}, 388-394.

\bibitem{CU7} Bazeia, D., Lobao, A. S., Losano, L., $\&$ Menezes, R. (2014). \textit{First-order formalism for twinlike models with several real scalar fields}. The European Physical Journal C, \textbf{74(2)}, 2755.

\bibitem{CU8} Mohammadi, M., $\&$ Riazi, N. (2019). \textit{The affective factors on the uncertainty in the collisions of the soliton solutions of the double field sine-Gordon system}. Communications in Nonlinear Science and Numerical Simulation, \textbf{72}, 176-193.
\bibitem{CU9} Riazi, N., Azizi, A., $\&$ Zebarjad, S. M. (2002). \textit{Soliton decay in a coupled system of scalar fields}. Physical Review D, \textbf{66(6)}, 065003.
\bibitem{CU10} Gani, V. A., $\&$ Kudryavtsev, A. E. (2001). \textit{Collisions of domain walls in a supersymmetric model}. Physics of Atomic Nuclei, \textbf{64(11)}, 2043-2050.
\bibitem{CU11} Gani, V. A., Ksenzov, V. G., $\&$ Kudryavtsev, A. E. (2010). \textit{Example of a self-consistent solution for a fermion on domain wall}. Physics of Atomic Nuclei, \textbf{73(11)}, 1889-1892.
\bibitem{CU12}  Gani, V. A., Ksenzov, V. G., $\&$ Kudryavtsev, A. E. (2011). \textit{Stable branches of a solution for a fermion on domain wall}. Physics of Atomic Nuclei, \textbf{74(5)}, 771-777.

\bibitem{CU13}  Gani, V. A., Konyukhova, N. B. , Kurochkin, S. V. , $\&$ Lensky V. A. (2004). \textit{Study of stability of a charged topological soliton in the system of two interacting scalar fields}.
computational mathematics and mathematical physics, \textbf{44}, 1968. 








\bibitem{CK} Mohammadi, M., $\&$ Riazi, N. (2014). \textit{Bi-dimensional soliton-like solutions of the nonlinear complex sine-Gordon system}. Progress of Theoretical and experimental Physics, \textbf{2014(2)}, 023A03.














\bibitem{EZ0} Rajaraman, R. (1982). \textit{Solitons and instantons}. North Holland, Elsevier, Amsterdam.

\bibitem{EZ00}   Campbell, D. K., Schonfeld, J. F., $\&$ Wingate, C. A. (1983). \textit{Resonance structure in kink-antikink interactions in $\varphi^4$ theory}. Physica D: Nonlinear Phenomena, \textbf{9(1-2)}, 1-32.


\bibitem{EZ1} Campbell, D. K., Peyrard, M., $\&$ Sodano, P. (1986). \textit{Kink-antikink interactions in the double sine-Gordon equation}. Physica D: Nonlinear Phenomena, \textbf{19(2)}, 165-205.

\bibitem{EZ2} Charkina, O. V., $\&$ Bogdan, M. M. (2006).\textit{ Internal modes of solitons and near-integrable highly-dispersive nonlinear systems}. Symmetry, integrability and geometry: methods and applications, \textbf{2(0)}, 47-12.

\bibitem{EZ3} Gharaati, A. R., Riazi, N., $\&$ Mohebbi, A. F. (2006). \textit{Internal modes of relativistic solitons}. \textit{International Journal of Theoretical Physics}, \textbf{45(1)}, 53-63.

\bibitem{EZ31} Morris, J. R. (2018). \textit{Small deformations of kinks and walls}. Annals of Physics, \textbf{393}, 122-131.

\bibitem{EZ4} Gani, V. A., $\&$ Kudryavtsev, A. E. (1999). \textit{Kink-antikink interactions in the double sine-Gordon equation and the problem of resonance frequencies}. Physical Review E, \textbf{60(3)}, 3305.

\bibitem{EZ5} Popov, C. A. (2005). \textit{Perturbation theory for the double sine-Gordon equation}. Wave Motion, \textbf{42(4)}, 309-316.

\bibitem{EZ6} Peyravi, M., Montakhab, A., Riazi, N., $\&$ Gharaati, A. (2009). \textit{Interaction properties of the periodic and step-like solutions of the double-Sine-Gordon equation}. The European Physical Journal B, \textbf{72(2)}, 269.

\bibitem{EZ7} Wazwaz, A. M. (2006). \textit{Compactons, solitons and periodic solutions for some forms of nonlinear Klein-Gordon equations}. Chaos, Solitons $\&$ Fractals, \textit{ 28(4)}, 1005-1013.


\bibitem{EZ9} Alonso-Izquierdo, A., $\&$ Guilarte, J. M. (2012). \textit{On a family of ($1+ 1$)-dimensional scalar field theory models: kinks, stability, one-loop mass shifts}. Annals of Physics, \textbf{327(9)}, 2251-2274.

\bibitem{EZ10} Gani, V. A., Kudryavtsev, A. E., $\&$ Lizunova, M. A. (2014). \textit{Kink interactions in the ($1+ 1$)-dimensional $\varphi^6$ model}. Physical Review D, \textbf{89(12)}, 125009

\bibitem{EZ11} Gani, V. A., Marjaneh, A. M., Askari, A., Belendryasova, E., $\&$ Saadatmand, D. (2018).\textit{ Scattering of the double sine-Gordon kinks}. The European Physical Journal C, \textbf{78(4)}, 345.

\bibitem{EZ12} Dorey, P., $\&$ Romańczukiewicz, T. (2018). \textit{Resonant kink–antikink scattering through quasinormal modes}. Physics Letters B, \textbf{779}, 117-123.

\bibitem{EZ13} Popov, S. P. (2014). \textit{Interactions of breathers and kink pairs of the double sine-Gordon equation}. Computational Mathematics and Mathematical Physics, \textbf{54(12)}, 1876-1885.

\bibitem{EZ14}   Bazeia, D., Gomes, A. R., Nobrega, K. Z., $\&$ Simas, F. C. (2019). \textit{Kink scattering in a hybrid model}. Physics Letters B, \textit{793}, 26-32.


\bibitem{EZ15} Khare, A., Christov, I. C., $\&$ Saxena, A. (2014). \textit{Successive phase transitions and kink solutions in $\phi^8$, $\phi^{10}$, and $\phi^{12}$ field theories}. Physical Review E, \textbf{90(2)}, 023208

\bibitem{EZ16} Bazeia, D., Belendryasova, E., $\&$ Gani, V. A. (2018). \textit{Scattering of kinks of the sinh-deformed $\varphi^ 4$ model}. The European Physical Journal C, \textbf{78(4)}, 340.
















\bibitem{IM1} Goatham, S. W., Mannering, L. E., Hann, R., $\&$ Krusch, S. (2011). \textit{Dynamics of Multi-kinks in the Presence of Wells and Barriers}.  Acta Physica Polonica B, \textbf{42},  2087.

\bibitem{IM2} Popov, S. P. (2013). \textit{Influence of dislocations on kink solutions of the double sine-Gordon equation}. Computational Mathematics and Mathematical Physics, \textbf{53(12)}, 1891-1899.

\bibitem{IM3} Saadatmand, D., Dmitriev, S. V., Borisov, D. I., $\&$ Kevrekidis, P. G. (2014). I\textit{interaction of sine-Gordon kinks and breathers with a parity-time-symmetric defect}. Physical Review E, \textbf{90(5)}, 052902.

\bibitem{IM4} Saadatmand, D., Dmitriev, S. V., Borisov, D. I., Kevrekidis, P. G., Fatykhov, M. A., $\&$ Javidan, K. (2015). \textit{Effect of the $\phi^4$ kink’s internal mode at scattering on a PT-symmetric defect}. JETP letters, \textbf{101(7)}, 497-502.

\bibitem{IM5} Saadatmand, D., Dmitriev, S. V., Borisov, D. I., Kevrekidis, P. G., Fatykhov, M. A., $\&$ Javidan, K. (2015). \textit{Kink scattering from a parity-time-symmetric defect in the $\phi^4$ model}. Communications in Nonlinear Science and Numerical Simulation, \textbf{29(1-3)}, 267-282.

\bibitem{IM6} Fei, Z., Kivshar, Y. S., $\&$ Vazquez, L. (1992). \textit{Resonant kink-impurity interactions in the $\varphi^4$ model}. Physical Review A, \textbf{46(8)}, 5214.

\bibitem{IM7} Fei, Z., Kivshar, Y. S., $\&$ Vazquez, L. (1992). \textit{Resonant kink-impurity interactions in the sine-Gordon model}. Physical Review A, \textbf{45(8)}, 6019.

\bibitem{IM8} Kivshar, Y. S., Fei, Z., $\&$ Vázquez, L. (1991). \textit{Resonant soliton-impurity interactions}. Physical review letters, \textbf{67(10)}, 1177.


\bibitem{IMPUr} Lizunova, M. A., Kager, J., de Lange, S., $\&$ van Wezel, J. (2020). \textit{ Kinks and realistic impurity models in $\varphi^ 4$-theory}. arXiv preprint arXiv:2007.04747.












\bibitem{PLT1} Christov, I. C., Decker, R. J., Demirkaya, A., Gani, V. A., Kevrekidis, P. G., $\&$ Radomskiy, R. V. (2019). \textit{Long-range interactions of kinks}. Physical Review D, \textbf{99(1)}, 016010.

\bibitem{PLT2} Belendryasova, E., $\&$ Gani, V. A. (2019). \textit{ Scattering of the $\varphi^8$ kinks with power-law asymptotics}. Communications in Nonlinear Science and Numerical Simulation, \textbf{67}, 414-426.

\bibitem{PLT3} Bazeia, D., Menezes, R., $\&$ Moreira, D. C. (2018). \textit{Analytical study of kinklike structures with polynomial tails}. Journal of Physics Communications, \textbf{ 2(5)}, 055019.

\bibitem{PLT4}  Christov, I. C., Decker, R. J., Demirkaya, A., Gani, V. A., Kevrekidis, P. G., Khare, A., $\&$ Saxena, A. (2019). \textit{Kink-kink and kink-antikink interactions with long-range tails}. Physical review letters, \textbf{122(17)}, 171601.

\bibitem{PLT5}  Manton, N. S. (2019). \textit{Forces between kinks and antikinks with long-range tails}. Journal of Physics A: Mathematical and Theoretical, \textbf{52(6)}, 065401.

\bibitem{PLT6} Christov, I. C., Decker, R. J., Demirkaya, A., Gani, V. A., Kevrekidis, P. G., $\&$ Saxena, A. (2020). \textit{Kink-Antikink Collisions and Multi-Bounce Resonance Windows in Higher-Order Field Theories}. arXiv preprint arXiv:2005.00154.


\bibitem{EV1} Gani, V. A., Moradi Marjaneh, A., $\&$ Saadatmand, D. (2019). \textit{Multi-kink scattering in the double sine-Gordon model}. The European Physical Journal C, \textbf{79(7)}, 620.

\bibitem{EV2} Moradi Marjaneh, A., Saadatmand, D., Zhou, K., Dmitriev, S. V., $\&$ Zomorrodian, M. E. (2017). \textit{High energy density in the collision of $N$ kinks in the $\phi^4$ model}. Communications in Nonlinear Science and Numerical Simulation, \textbf{49}, 30-38.

\bibitem{EV3} Moradi Marjaneh, A., Gani, V. A., Saadatmand, D., Dmitriev, S. V., $\&$ Javidan, K. (2017). \textit{Multi-kink collisions in the $\phi^6$ model}. Journal of High Energy Physics, \textbf{2017(7)}, 28.

\bibitem{CC0} Hassanabadi, H., Lu, L., Maghsoodi, E., Liu, G., $\&$ Zarrinkamar, S. (2014). \textit{Scattering of Klein-Gordon particles by a Kink-like potential}. Annals of Physics, \textbf{342}, 264-269.

\bibitem{CC1} Takyi, I., $\&$ Weigel, H. (2016). \textit{Collective coordinates in one-dimensional soliton models revisited}. Physical Review D, \textbf{94(8)}, 085008.

\bibitem{CC2} Baron, H. E., Luchini, G., $\&$ Zakrzewski, W. J. (2014). \textit{Collective coordinate approximation to the scattering of solitons in the ($1+1$) dimensional NLS model}. Journal of Physics A: Mathematical and Theoretical, \textbf{47(26)}, 265201.

\bibitem{CC3} Javidan, K. (2010). \textit{Collective coordinate variable for soliton-potential system in sine-Gordon model}. Journal of mathematical physics, \textbf{51(11)}, 112902.

\bibitem{CC4} Christov, I., $\&$ Christov, C. I. (2008). \textit{Physical dynamics of quasi-particles in nonlinear wave equations}. Physics Letters A, \textbf{372(6)}, 841-848

\bibitem{Man3} Radomskiy, R. V., Mrozovskaya, E. V., Gani, V. A., $\&$ Christov, I. C. (2017). \emph{Topological defects with power-law tails}. In J. Phys. Conf. Ser, \textbf{798},  012087.


\bibitem{Man1} Manton, N. S. (1979). \textit{An effective Lagrangian for solitons}. Nuclear Physics B, \textbf{150}, 397-412.

\bibitem{Man2} Kevrekidis, P. G., Khare, A., $\&$ Saxena, A. (2004). \textit{Solitary wave interactions in dispersive equations using Manton’s approach}. Physical Review E, \textbf{70(5)}, 057603.

\bibitem{INTER1} Peyrard, M., $\&$ Campbell, D. K. (1983). \textit{Kink-antikink interactions in a modified sine-Gordon model}. Physica D: Nonlinear Phenomena, \textbf{9(1-2)}, 33-51.

\bibitem{FR2}  Goodman, R. H., $\&$ Haberman, R. (2005). \textit{Kink-Antikink Collisions in the $\phi^4$ Equation: The $n$-Bounce Resonance and the Separatrix Map}. SIAM Journal on Applied Dynamical Systems, \textbf{4(4)}, 1195-1228.

\bibitem{INTER2} Charkina, O. V., $\&$ Bogdan, M. M. (2006). \textit{Internal modes of solitons and near-integrable highly-dispersive nonlinear systems}. Symmetry, integrability and geometry: methods and applications, \textbf{2(0)}, 47-12.

\bibitem{AI1} Kivshar, Y. S., Pelinovsky, D. E., Cretegny, T., $\&$ Peyrard, M. (1998). I\textit{nternal modes of solitary waves}. Physical review letters, \textbf{80(23)}, 5032.

\bibitem{AI2} Mohammadi, M., $\&$ Riazi, N. (2011). \textit{Approaching integrability in bi-dimensional nonlinear field equations}. Progress of Theoretical Physics, \textbf{126(2)}, 237-248.

\bibitem{RD}   Mohammadi, M., Riazi, N., $\&$ Azizi, A. (2012). \textit{Radiative Properties of Kinks in the $\sin^4(\phi)$ System}. Progress of theoretical physics, \textbf{128(4)}, 615-627.

\bibitem{TV1} Campbell, D. K., $\&$ Peyrard, M. (1986). \textit{Solitary wave collisions revisited}. Physica D: Nonlinear Phenomena, \textbf{18(1-3)}, 47-53.

\bibitem{TV2} Belova, T. I., $\&$ Kudryavtsev, A. E. (1997). \textit{Solitons and their interactions in classical field theory}. Physics-Uspekhi, \textbf{40(4)}, 359.

\bibitem{FR1} Anninos, P., Oliveira, S., $\&$ Matzner, R. A. (1991). \textit{Fractal structure in the scalar $\lambda (\varphi^2-1)^2$ theory}. Physical Review D, \textbf{44(4)}, 1147.

\bibitem{FR3} Goodman, R. H., $\&$ Haberman, R. (2007). \textit{Chaotic scattering and the n-bounce resonance in solitary-wave interactions}. Physical review letters, \textbf{98(10)}, 104103.

\bibitem{FR4}  Gani, V. A., Lensky, V., $\&$ Lizunova, M. A. (2015). \textit{Kink excitation spectra in the ($1+ 1$)-dimensional $\varphi^8$ model}. Journal of High Energy Physics, \textbf{2015(8)}, 147.

\bibitem{PHI41} Quintero, N. R., Sánchez, A., $\&$ Mertens, F. G. (2000). \textit{Resonances in the dynamics of $\varphi^4$ kinks perturbed by ac forces}. Physical Review E, \textbf{62(4)}, 5695.

\bibitem{newg} Alonso-Izquierdo, A., Nieto, L. M., $\&$ Queiroga-Nunes, J. (2020). \textit{Scattering between wobbling kinks}. arXiv preprint arXiv:2007.15517.

\bibitem{PHI42} Fei, Z., Konotop, V. V., Peyrard, M., $\&$ Vázquez, L. (1993). \textit{Kink dynamics in the periodically modulated $\varphi^4$ model}. Physical Review E, \textbf{48(1)}, 548.























































%







































\bibitem{Azizi} Azizi, A., $\&$ Mohammadi, M. (2010). \textit{Separability in 1+ 1 Dimensions in Classical Nonlinear Fields}. arXiv preprint arXiv:1004.4780.

\bibitem{Derrick} Derrick, G. H. (1964). \textit{Comments on nonlinear wave equations as models for elementary particles}. Journal of Mathematical Physics, \textbf{5(9)}, 1252-1254

\bibitem{Mohammadi:2020yku}, M.~Mohammadi and R.~Dehghani,
\textit{Kink-antikink collisions in the periodic $\varphi^{4}$ model,}
Commun. Nonlinear Sci. Numer. Simul. \textbf{94} (2021), 105575
doi:10.1016/j.cnsns.2020.105575
[arXiv:2005.11398 [nlin.PS]].

\end{thebibliography}
\end{document}